\documentclass{aa}  

\usepackage{graphicx}
\usepackage{txfonts}
\usepackage{amsmath}
\usepackage{hyperref}
\usepackage{cleveref}
\usepackage{xcolor}

\crefname{figure}{Fig.}{Figs.}
\Crefname{figure}{Figure}{Figures}

\crefname{equation}{Eq.}{Eqs.}
\Crefname{equation}{Equation}{Equations}

\crefname{section}{Sect.}{Sects.}
\Crefname{section}{Section}{Sections}

\crefname{appendix}{App.}{Apps.}
\Crefname{appendix}{Appendix}{Appendices}

\newcommand{\ha}{\ensuremath{\mathrm{H}\alpha}}
\newcommand{\hb}{\ensuremath{\mathrm{H}\beta}}
\newcommand{\oiii}{\ensuremath{[\ion{O}{III}]}}
\newcommand{\oiiihb}{\ensuremath{\oiii{}/\hb{}}}

\newcommand{\lumha}{\ensuremath{L_{\mathrm{H}\alpha}}}
\newcommand{\lumhaint}{\ensuremath{L_{\mathrm{H}\alpha}^{\mathrm{int}}}}
\newcommand{\lumhaobs}{\ensuremath{L_{\mathrm{H}\alpha}^{\mathrm{obs}}}}
\newcommand{\lumhbobs}{\ensuremath{L_{\mathrm{H}\beta}^{\mathrm{obs}}}}

\newcommand{\loglumhaint}{\ensuremath{\log_{10} \left( L_{\mathrm{H}\alpha}^{\mathrm{int}} / \mathrm{erg}\,\mathrm{s}^{-1}\right)}}
\newcommand{\loglumhafracint}{\ensuremath{\log_{10} \left[ \frac{L_{\mathrm{H}\alpha}^{\mathrm{int}}}{\mathrm{erg}\,\mathrm{s}^{-1}}\right]}}
\newcommand{\loglumhafracobs}{\ensuremath{\log_{10} \left[ \frac{L_{\mathrm{H}\alpha}^{\mathrm{obs}}}{\mathrm{erg}\,\mathrm{s}^{-1}}\right]}}

\newcommand{\ewha}{\ensuremath{\mathrm{EW}_{\mathrm{H}\alpha}}}
\newcommand{\ewhaobs}{\ensuremath{\mathrm{EW}_{\mathrm{H}\alpha}^{\mathrm{obs}}}}

\newcommand{\logewhafracint}{\ensuremath{\log_{10} \left[ \frac{\mathrm{EW}_{\mathrm{H}\alpha}^{\mathrm{int}}} {\AA} \right]}}
\newcommand{\logewhafracobs}{\ensuremath{\log_{10} \left[ \frac{\mathrm{EW}_{\mathrm{H}\alpha}^{\mathrm{obs}}} {\AA} \right]}}

\newcommand{\bd}{\ensuremath{\ha{}/\hb{}}}
\newcommand{\logbd}{\ensuremath{\log_{10} \left( \lumhaobs{} / \lumhbobs{} \right)}}
\newcommand{\aha}{\ensuremath{A_{\mathrm{H}\alpha}}}

\newcommand{\sfrvsha}{SFR---\lumha{}}
\newcommand{\sfrvsmstar}{SFR---$M_*$}
\newcommand{\sfrmodel}{SFR(\ha{})}
\newcommand{\cha}{\ensuremath{C_{\mathrm{H}\alpha}}}
\newcommand{\sfrvar}{\ensuremath{\sigma_{\mathrm{SFR}}}}
\newcommand{\sfrbias}{\ensuremath{\Delta_{\mathrm{SFR}}}}
\newcommand{\sfrbiasvar}{\ensuremath{\sigma_{\Delta_{\mathrm{SFR}}}}}
\newcommand{\logsfrfrac}{\ensuremath{\log_{10} \left[ \frac{\mathrm{SFR}}{M_\odot\,\mathrm{yr}^{-1}}\right]}}
\newcommand{\sfrd}{\ensuremath{\rho_{\mathrm{SFR}}}}
\newcommand{\sfms}{SFMS}
\newcommand{\tsfr}{\ensuremath{t_{\mathrm{SFR}}}}
\newcommand{\tage}{\ensuremath{t_{\mathrm{age}}}}

\newcommand{\jwst}{\textit{JWST}}

\newcommand{\sphinx}{SPHINX}
\newcommand{\sphinxtwenty}{\sphinx{}$^{20}$}
\newcommand{\bpass}{\textsc{BPASSv$2.2$}}
\newcommand{\cloudy}{\textsc{CLOUDY}}
\newcommand{\rascas}{\textsc{RASCAS}}
\newcommand{\ramsesrt}{\textsc{RAMSES-RT}}

\newcommand{\pcrit}{\ensuremath{10^{-6}}}

\begin{document}

    \title{\ha{} as a tracer of star formation in the SPHINX cosmological simulations}
    \titlerunning{\ha{} as a tracer of star formation in \sphinx{}}

    \author{
        I.~G.~Kramarenko \inst{1}\fnmsep\thanks{email: ivan.kramarenko@ista.ac.at},
        J.~Rosdahl \inst{2},
        J.~Blaizot \inst{2},
        J.~Matthee \inst{1},
        H.~Katz \inst{3}\fnmsep\inst{4}
        \and
        C.~Di Cesare \inst{1}
    }

    \authorrunning{Kramarenko et al.}

    \institute{
        Institute of Science and Technology Austria (ISTA), Am Campus 1, 3400 Klosterneuburg, Austria
        \and
        Univ Lyon, Univ Lyon1, Ens de Lyon, CNRS, Centre de Recherche Astrophysique de Lyon UMR5574, F-69230, Saint-Genis-Laval, France
        \and
        Department of Astronomy \& Astrophysics, University of Chicago, 5640 S Ellis Avenue, Chicago, IL 60637, USA
        \and
        Kavli Institute for Cosmological Physics, University of Chicago, Chicago IL 60637, USA
    }
    
    \date{Received XXX; accepted YYY}
 
  \abstract{
  The \ha{} emission line in galaxies is a powerful tracer of their recent star formation activity. With the advent of \jwst{}, we are now able to routinely observe \ha{} in galaxies at high redshift ($z \gtrsim 3$) and thus measure their star formation rates (SFRs). However, using classical \sfrmodel{} calibrations to derive the SFRs leads to biased results because high-redshift galaxies are commonly characterized by low metallicities and bursty star formation histories, affecting the conversion factor between the \ha{} luminosity (\lumha{}) and the SFR. In this work, we develop a set of new \sfrmodel{} calibrations that allow us to predict the SFRs of \ha{}-emitters at $z \gtrsim 3$ with minimal error. We use the \sphinx{} cosmological simulations to select a sample of star-forming galaxies representative of the \ha{}-emitter population observed with \jwst{}.
  We then derive linear corrections to the classical \sfrmodel{} calibrations, taking into account variations in the physical properties (e.g., stellar metallicities) among individual galaxies. We obtain two new \sfrmodel{} calibrations that, compared to the classical calibrations, reduce the root mean squared error (RMSE) in the predicted SFRs by $\Delta \mathrm{RMSE} \approx 0.04$~dex and $\Delta \mathrm{RMSE} \approx 0.06$~dex, respectively. Using the recent \jwst{} NIRCam/grism observations of \ha{}-emitters at $z \sim 6$, we show that the new calibrations affect the high-redshift galaxy population statistics: (i) the estimated cosmic star formation rate density decreases by $\Delta \rho_{\mathrm{SFR}}\approx 12$\%, and (ii) the observed slope of the star formation main sequence increases by $\Delta \partial \log \mathrm{SFR} / \partial \log \mathrm{M_\star} = 0.08 \pm 0.02$.
  }

   \keywords{galaxies: fundamental parameters -- galaxies: evolution -- galaxies: formation -- galaxies: high-redshift -- galaxies: star formation}

   \maketitle

\section{Introduction}
One of the most fundamental and basic properties of galaxies is the rate at which new stars form, the star formation rate (SFR). The SFR is sensitive to the interplay between the gas inflow and outflow rates of galaxies, and the efficiency at which this gas is converted into stars \citep[e.g.,][]{Bouche10,Lilly13,Behroozi19}. The SFR can fluctuate on different timescales \citep[e.g.,][]{Sparre15,Iyer2020,Flores21}, which range from dynamical timescales of molecular clouds ($\sim10$s of Myrs), through galaxy merger timescales ($\sim100$~Myr), to long (Gyr-scale) coherences in accretion rates \citep[e.g.,][]{MattheeSchaye19,EWang20,Tacchella20,Wan25}. Observationally, SFRs can be measured using various diagnostics \citep[e.g.,][]{Kennicutt2012,DeLooze2014,Davies19} that are either based on the light from massive, short-lived stars (e.g., the rest-frame UV), or processed radiation (e.g., nebular lines in HII regions or dust emission). Each of these diagnostics probes different timescales of star formation and is affected by dust attenuation differently.

At high redshifts ($z>3$), the rest-frame UV emission used to be the most sensitive SFR indicator that could be measured for large galaxy samples, tracing star formation on $t \sim 100$~Myr timescales \citep[e.g.,][]{Bouwens15}. However, dust corrections in the rest-frame UV are significant, with potentially up to 50\% of the cosmic star formation rate density (\sfrd{}) at $z\sim5$ being obscured by dust \citep[e.g.,][]{Zavala21,Sun25}. This emphasizes the great importance of alternative SFR indicators that are less sensitive to dust attenuation, such as the \ha{} emission. The \ha{} photons are emitted in the rest-frame optical ($\lambda = 6564$~{\AA}), where dust corrections are smaller than in the rest-frame UV due to the negative slope of the dust attenuation curve (however, we caution that there is growing evidence of flatter attenuation curves at fixed $A_V$ or stellar mass at high redshifts compared to the local Universe; e.g., \citealp{Shivaei2025,Fisher2025}). The luminosity of the \ha{} line is directly proportional to the luminosity of ionizing photons that are predominantly emitted by very young stellar populations, making \ha{} sensitive to star formation on shorter timescales \citep[$t \sim 10$~Myr; e.g.,][]{Leitherer99}. As the \jwst{} can now measure the distribution of \ha{} luminosities out to $z\sim7$ based on spectroscopy and narrow-band imaging \citep[e.g.,][]{Clarke24,Pirie24,Covelo-Paz25,Fu25}, it is timely to reassess the role of the H$\alpha$ emission as a tracer of the SFR.

The conversion factor between the H$\alpha$ luminosity\footnote{Throughout this work, we assume a negligible contribution of active galactic nuclei to the H$\alpha$ emission.} and the SFR primarily depends on assumptions on the properties of stellar populations, in particular, the star formation history (SFH), the initial mass function (IMF), binarity, and stellar metallicity \citep[e.g,][]{Wilkins2019}. The SFH and IMF are relevant because the ionizing photon emissivity from a single stellar population drops sharply after a few Myr when massive, short-lived O stars leave the main sequence. Binary interactions can extend these ionizing lifetimes \citep{Gotberg17}, smoothing the otherwise steep drop-off after a few Myrs. At fixed stellar mass and age, the metallicity affects the effective temperature of stellar atmospheres due to line blanketing effects, affecting the ionizing luminosity as well.

The classical \sfrmodel{} calibration proposed by \cite{Kennicutt1998} (see also \citealt{Kennicutt2012}) assumes a constant SFH over $t=100$~Myr and solar metallicity, based on the typical conditions in galaxies in the low-redshift Universe. However, various spectroscopic observations of galaxies at high redshifts suggest that their SFHs are bursty \citep[e.g.,][]{Asada24,Endsley24,Cole25,Carvajal-Bohorquez2025}. Further, given the gas-phase oxygen abundances are typically measured to be $\sim10$\% solar \citep{Curti24,Scholte25}, and early galaxies are expected to be $\alpha$-enhanced due to the rapid formation timescales \citep[e.g.,][]{Chruslinska24}, the iron-peak element abundances are likely to be low. This implies hotter stars and a more efficient \ha{} photon production at a fixed SFR \citep[e.g.,][]{Theios2019} than at low redshift.

The aim of this paper is to revisit the classical \sfrmodel{} calibrations to predict the SFRs of galaxies at high redshift ($z \gtrsim 3$). Instead of using idealistic models with constant SFHs and fixed stellar metallicities, we use simulated galaxies and their modeled \ha{} luminosities from the \sphinx{} cosmological radiation-hydrodynamical simulation \citep{Rosdahl2018,Katz2023}. These cosmological simulations self-consistently model SFHs and metal enrichment processes and are calibrated to match various observables at high redshifts ($z \sim 6$). The simulated volume matches the typical volumes of deep JWST surveys well, which enables a high resolution that can capture SFR variations on short timescales ($t \sim 1$~Myr) and approximate the distribution of gas in the interstellar medium relevant for radiative transfer effects.

In \cref{sec:methods}, we briefly describe the \sphinx{} simulations and explain how the H$\alpha$ luminosities are derived. In \cref{sec:results}, we present a set of simulation-inferred \sfrmodel{} calibrations that predict the SFRs of \sphinx{} galaxies with minimal error. Finally, we discuss the implications of our work in \cref{sec:discussion} and summarize in \cref{sec:summary}.

\section{Methods}
\label{sec:methods}

In this work, we use the simulated galaxy data from \sphinx{}, a suite of cosmological radiation-hydrodynamic simulations \citep{Rosdahl2018}. \sphinx{} leverages the adaptive mesh refinement code \ramsesrt{} \citep{Rosdahl2013,Teyssier2002} to simulate tens of thousands of galaxies in the epoch of reionization. The ISM is resolved down to $76$~co-moving~pc, corresponding to a physical resolution of $11$~pc at $z=6$. The dark matter halos are resolved down to the atomic cooling threshold ($M_{\mathrm{vir}}\approx3\times10^7$~M$_{\odot}$). Hydrodynamics, non-equilibrium thermochemistry, and radiative transfer of Lyman-continuum radiation (LyC; $\lambda < 912$\r{A}) are performed self-consistently on the fly, with subgrid implementations for gas cooling, star formation, metal enrichment, and feedback. The on-the-fly radiative emission and propagation of LyC photons make \sphinx{} an excellent model for accurately predicting how \ha{} emission traces star formation in the simulated galaxies.

Our analysis is based on the data from the \sphinxtwenty{} Public Data Release, Version~1 \citep[SPDRv1;][]{Katz2023}. \sphinxtwenty{} is the largest simulation of the \sphinx{} suite, with a periodic box with volume of ($20$~cMpc)$^3$ simulated down to a redshift of $z=4.64$ \citep{Rosdahl2022}. \sphinxtwenty{} uses the \bpass{} stellar population synthesis (SPS) model \citep{Stanway18}, which adopts a \cite{Kroupa2001} IMF with an upper mass cutoff of 100~M$_{\odot}$ and a slope of $-1.3$ from $0.1$ to $0.5$~M$_{\odot}$ and $-2.35$ from $0.5$ to $100$~M$_{\odot}$. The intrinsic Balmer line luminosities are generated by interpolating the \cloudy{} v17.03 models \citep{Ferland2017} for gas cells that host stellar particles with unresolved Stromgren spheres, or by using the non-equilibrium ionizing fractions from the simulation otherwise. For each halo, the total intrinsic line emission is then the sum of the line luminosities of all gas cells within the virial radius of the halo. The equivalent widths are calculated by taking the stellar and nebular continuum emission into account, generated as detailed in \citet{Katz2023}.

We post-process the line and continuum radiation, subject to absorption and scattering by dust, using the radiative transfer code \rascas{} \citep{Michel-Dansac2020}. Dust is assigned to the gas cells using the Small Magellanic Cloud (SMC) dust model from \citet{Laursen2009}. We note that the dust attenuation curve at high redshifts likely has a steeper slope than the SMC slope \citep[e.g.,][]{Shivaei2025,Markov2025}, possibly due to changes in the dust grain properties. Next, 200,000 photon packets are probabilistically distributed to the gas cells, with the initial wavelengths placed at the line centers, thermally broadened, and shifted according to the bulk velocity of the cell. Likewise, 200,000 photon packets are probabilistically distributed to the star particles (gas cells) based on the luminosities of the stellar (nebular) continuum at the line centers. The observed line luminosities and equivalent widths are then obtained by computing the escape fractions of each component (for details, see \citealp{Katz2023}). 

Due to the computation time and data storage limitations, SPDRv1 only includes \sphinxtwenty{} galaxies with $\mathrm{SFR}_{10} \geq 0.3$~M$_{\odot}$~yr$^{-1}$, where $\mathrm{SFR}_{10}$ is the SFR averaged over $t=10$~Myr. This excludes objects that are likely too faint to be observed with \jwst{} \citep{Katz2023}, unless lensing is considered \citep[e.g.,][]{Bezanson2024,Naidu24}. However, an SFR cut also leads to an incompleteness in the \sfrvsha{} relation at low luminosities, which can bias our results. We therefore limit our selection of galaxies from SPDRv1 to those with an observed \ha{} luminosity (\lumhaobs{}) greater than $\lumhaobs{} = 1.8 \times 10^{41}$~erg~s$^{-1}$ (here, we use the simulation-based luminosities calculated as described above). This value is motivated by the recent \ha{} LF measurements from \jwst{} NIRCam/grism, which reach similar luminosities at the faint end \citep[e.g.,][]{Fu25}. Crucially, this selection yields a complete \ha{}-emitter sample, which we check by running \rascas{} on all \sphinxtwenty{} halos whose intrinsic \ha{} luminosity (\lumhaint{}) exceeds the luminosity cut. We also try $\lumhaobs{} > 1.0 \times 10^{41}$~erg~s$^{-1}$ and $\lumhaobs{} > 3.0 \times 10^{41}$~erg~s$^{-1}$ cuts, finding little impact on our results. With this additional luminosity cut, our final sample comprises $N=63$, 66, 48, 36, 22, 17, and 13 galaxies at $z = 4.64$, 5, 6, 7, 8, 9, and 10, respectively, with a total of $N=265$ galaxies.

The SN feedback model in \sphinx{} is calibrated to match the UV luminosity function (UVLF) at $z = 6$ (see App.~C in \citealp{Rosdahl2018}). A possible caveat is that the stellar mass-halo mass relation in \sphinx{} is $0.5$--$1.0$~dex higher than predicted by the \citet{Behroozi19} model, for example, which matches a broad array of observational data, including stellar mass functions at $z=0$--$4$ and median UV---stellar mass relations at $z=4$--$8$ (see Fig.~4 in \citealp{Katz2023}). Furthermore, the stellar mass function (SMF) in \sphinx{} is significantly overestimated compared to the SMF inferred from the \jwst{}/NIRCam observations at $z=5$--$6$ \citep{Weibel2024}. If the stellar masses of \sphinx{} galaxies are overestimated, the SFRs should also be overestimated on average. This implies that the amount of dust in the simulation should be higher than in the real Universe in order to match the UVLF. The excess of dust would then explain the fact that \sphinx{} predicts lower \ha{} and \oiii{} luminosities at fixed stellar mass than those measured by \jwst{} \citep[e.g.,][]{Covelo-Paz25,Meyer2024}. Although the possibility of \sphinx{} galaxies having too much dust does not affect our main results as they are based on the intrinsic \ha{} properties (\cref{sec:results}), it will become important later for understanding the errors in the \sfrmodel{} calibrations that use the observed \ha{} (\cref{sec:discussion}).

\section{Results}
\label{sec:results}

In this section, we investigate the intrinsic (i.e., without taking dust attenuation into account) \sfrvsha{} relation in the \sphinxtwenty{} cosmological simulation. We compare this relation to the \sfrmodel{} calibrations from the literature and study the effect of the variations in the physical properties of individual galaxies (metallicity, stellar age, etc.) on the accuracy of the predicted SFRs. Based on this analysis, we develop a set of new \sfrmodel{} calibrations that minimize the error in \ha{}-derived SFRs at high redshift.

Unless otherwise specified, we use the SFRs averaged over the last $t = 10$~Myr of the SFH. This choice is motivated by the fact that nebular emission lines such as \ha{} trace stars with lifetimes of $t\sim 3$--10~Myr \citep[e.g.,][]{Kennicutt2012}. This is further supported by the simulation itself, where the correlation coefficient between the SFR and \lumhaint{} peaks between $4$~$\mathrm{Myr}\lesssim t \lesssim 12$~Myr ($r \approx 0.95$--$0.98$; see \cref{sec:sf-timescales}).

\subsection{The \sfrvsha{} relation in \sphinx{}}

\begin{figure}
    \centering
    \includegraphics[width=\linewidth]{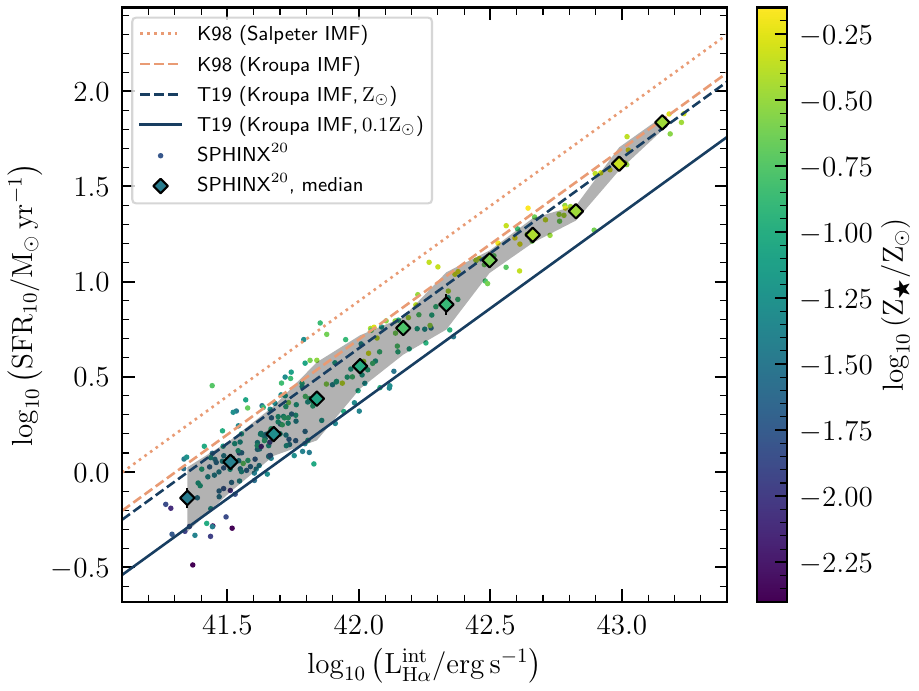}
    \caption{Intrinsic \sfrvsha{} relation in the \sphinxtwenty{} cosmological simulation, color-coded by the stellar metallicity. The SFR values are averaged over $t=10$~Myr, i.e., the typical lifetime of stars traced by \ha{}. The diamonds show the median SFR in the bins of \lumhaint{}, with the error bars showing the standard error on the median. The shaded region indicates the 68\% ($1\sigma$) confidence interval. The diagonal lines show the \sfrmodel{} calibrations from the literature, i.e. (top to bottom): the original K98 calibration, the K98 calibration converted to the Kroupa IMF, and two calibrations from T19 calculated for the Kroupa IMF and two different metallicities ($Z_*=Z_\odot$ and $Z_*=0.1Z_\odot$, respectively).}
    \label{fig:sfr-vs-ha-int}
\end{figure}

\Cref{fig:sfr-vs-ha-int} shows the intrinsic \sfrvsha{} relation for the galaxy sample selected from the \sphinxtwenty{} simulation as described in \cref{sec:methods}. In the same figure, we show different \sfrmodel{} calibrations proposed in the literature. These include the original K98 calibration ($\cha{} = -41.1$), the K98 calibration converted to the Kroupa IMF (\citealp{Kroupa2001}; $\cha{} = -41.3$), and two calibrations from \citet[][hereafter T19]{Theios2019} calculated for a Kroupa-type IMF and solar metallicity ($\cha{} = -41.35$), and the same IMF but subsolar metallicity ($Z_*=0.1Z_\odot$; $\cha{} = -41.64$), respectively. Here, \cha{} is the conversion factor between \lumhaint{} and SFR, defined as follows:
\begin{equation}
    \label{eq:sfr-calib-lit}
    \logsfrfrac{} = \loglumhafracint{} + \cha{}.
\end{equation}

\Cref{fig:sfr-vs-ha-int} reveals that a constant \cha{} cannot provide an accurate estimate of the SFRs of \sphinx{} galaxies across the full \lumhaint{} range. Specifically, the T19 calibration that assumes $Z_*=Z_\odot$ (the dashed blue line in \cref{fig:sfr-vs-ha-int}) matches the median \sfrvsha{} relation in \sphinx{} at high luminosities well ($\loglumhaint{} \gtrsim 42.2$). In contrast, the same \sfrmodel{} calibration at low luminosities ($\loglumhaint{} \lesssim 42.2$) overestimates the SFRs by $\approx 0.1$--$0.2$~dex. Decreasing \cha{} would result in a better match at low luminosities, but the SFRs at high luminosities would then be underestimated.

We note that the median \sfrvsha{} relation in \sphinx{} at low luminosities lies between the two T19 calibrations that differ only in the assumption on metallicity ($Z_*=0.1Z_\odot$ vs.~$Z_*=Z_\odot$; indicated by the solid and dashed blue lines in \cref{fig:sfr-vs-ha-int}, respectively). Motivated by this result, we test whether the variations in metallicity might be a key factor driving the evolution of the median \sfrvsha{} relation in \sphinx{}. We split our sample into low- and high-\lumhaint{} groups ($\loglumhaint{} < 42.2$ and $\loglumhaint{} > 42.2$, respectively), and calculate the median stellar metallicity\footnote{Hereafter, stellar metallicities (and ages) of \sphinx{} galaxies are weighted by the LyC luminosity.} in each group. We obtain $Z_*=0.06 Z_\odot$ and $Z_*=0.33 Z_\odot$ in the low- and high-\lumhaint{} groups, respectively. In other words, brighter galaxies are more metal-rich on average, which reflects the mass-metallicity relation (MZR) in \sphinx{} (for a discussion, see \citealp{Katz2023}), and likely causes an upturn of the median \sfrvsha{} relation at high luminosities (see also \cref{fig:sfr-vs-ha-int}, where the markers are color-coded by metallicity).

\Cref{fig:sfr-vs-ha-int} further demonstrates that the original K98 calibration (indicated by the dotted orange line) overestimates the SFRs by $\approx0.3$--$0.5$~dex compared to \sphinx{}. One of the main reasons for this discrepancy is the difference in the IMFs: K98 uses the \citet{Salpeter1955} IMF, whereas \sphinx{} uses a Kroupa-type IMF. After rescaling to the Kroupa IMF, the K98 calibration still predicts $\approx0.1$--$0.3$~dex higher SFRs than the simulation (the dashed orange line in \cref{fig:sfr-vs-ha-int}). This difference cannot be eliminated even when using the T19 calibration which, unlike the K98 calibration, assumes the same stellar population synthesis model as \sphinx{} (i.e., \bpass{}; the dashed blue line in \cref{fig:sfr-vs-ha-int}). Therefore, the only way to fully reconcile the K98 calibration and the simulation data is to assume a sub-solar metallicity, for some galaxies as low as $Z_* = 0.1 Z_\odot$ (the solid blue line in \cref{fig:sfr-vs-ha-int}). As discussed in the introduction, a subsolar metallicity like this agrees well with the current estimates of the gas-phase metallicity and also with the direct estimates of stellar metallicities of galaxies at $z\sim2-3$ \citep{Steidel2016,Cullen19}, although we caution that these were derived assuming constant SFHs, which might impact the results \citep{Matthee22}. 

\Cref{fig:sfr-vs-ha-int} also reveals that the \sfrvsha{} relation in \sphinx{} exhibits significant scatter around the median relation (hereafter, \sfrvar{}). In particular, we find that \sfrvar{} decreases from $\sfrvar{}\approx0.17$~dex at $\loglumhaint{}=41.3$ to $\sfrvar{}\approx0.04$~dex at $\loglumhaint{}=42.8$, and has a median $\sfrvar = 0.11$~dex. This scatter is neglected by the existing \sfrmodel{} calibrations, leading to errors in the predicted SFRs of individual galaxies. To quantify this effect, we introduce the SFR bias factor, defined as the ratio of the SFR predicted from \ha{} to the true SFR directly tracked by the simulation: $\sfrbias{} \equiv \mathrm{SFR}\left(\mathrm{H}\alpha\right)/\mathrm{SFR}_{10}$. Here, we use the T19 calibration that assumes $Z_*=Z_\odot$ to calculate \sfrmodel{} as this calibration provides the best match to the SPHINX data (see \cref{fig:sfr-vs-ha-int}). We note that using a different calibration would only change \sfrbias{} by a constant factor (in logarithmic scale) without affecting any trends observed between \sfrbias{} and the physical properties of \sphinx{} galaxies.

\begin{figure}
    \centering
    \includegraphics[width=\linewidth]{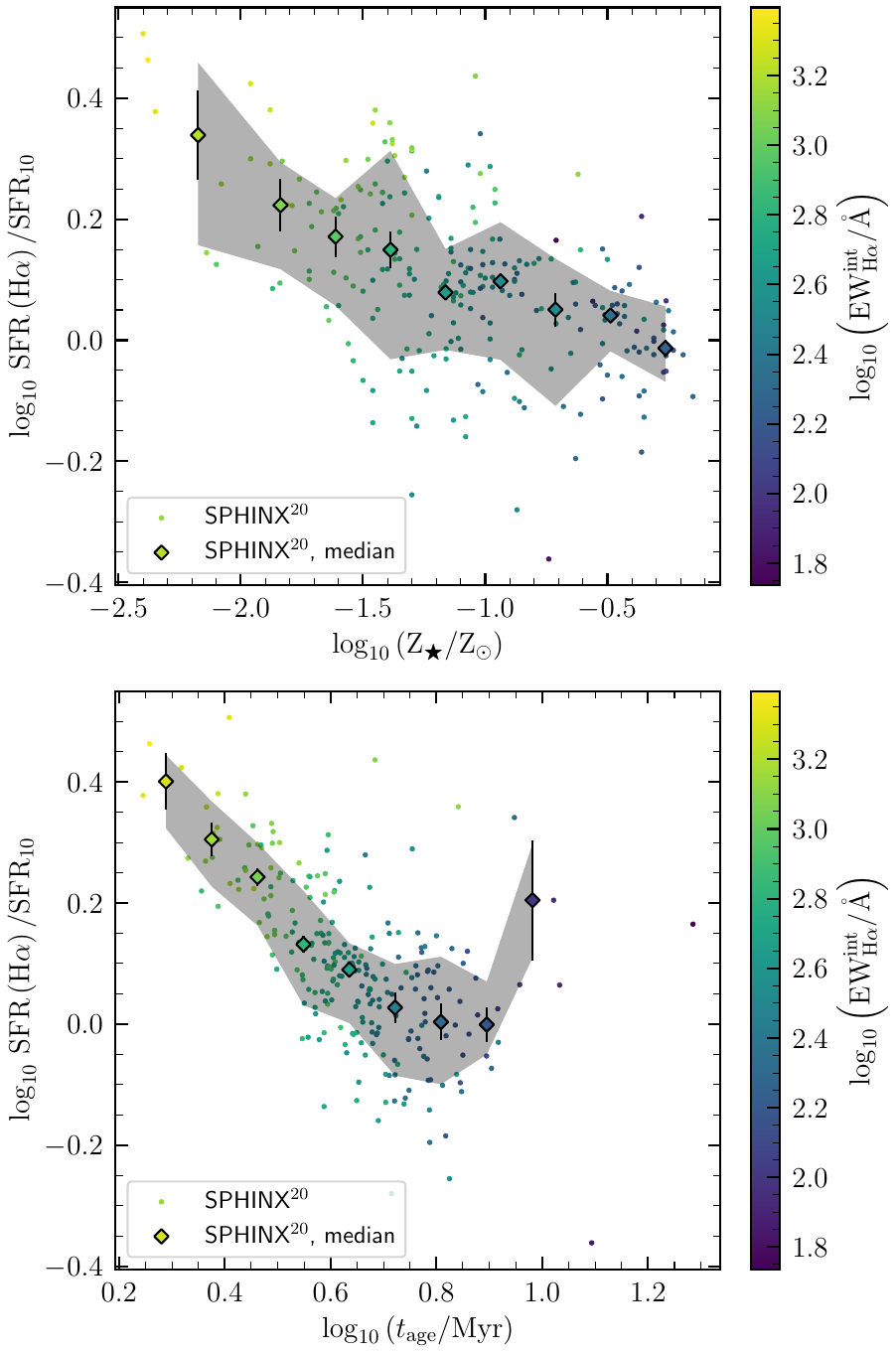}
    \caption{SFR bias factor ($\sfrbias{} \equiv \mathrm{SFR}\left(\mathrm{H}\alpha\right)/\mathrm{SFR}_{10}$) as a function of stellar metallicity (top) and age (bottom) in \sphinx{}, color-coded by \ewha{}. The \sfrmodel{} values are calculated using the T19 calibration ($Z_*=Z_\odot$). The diamonds show the running median, with the error bars indicating the standard error on the median. The shaded region indicates the 68\% ($1\sigma$) confidence interval.}
    \label{fig:dsfr-vs-star-prop}
\end{figure}

We find that $\sfrbias{}$ decreases with stellar metallicity ($r=-0.51$, $p<\pcrit{}$; \cref{fig:dsfr-vs-star-prop}, top panel), in agreement with our earlier analysis of the median \sfrvsha{} relation in \sphinx{}. However, we also report a significant scatter in \sfrbias{} at fixed $Z_*$: $ \sigma \approx 0.09$~dex. This indicates that metallicity alone does not cause the variations in the SFRs of \sphinx{} galaxies with respect to the SFRs predicted from \ha{}. In particular, we find that $\sfrbias{}$ shows a negative correlation with stellar age ($r=-0.59$, $p<\pcrit{}$; \cref{fig:dsfr-vs-star-prop}, bottom panel), with an average scatter of $\sigma \approx 0.07$~dex. Notably, galaxies with young stellar populations ($\tage{}\lesssim5$~Myr) exhibit high $\sfrbias{}$ ($\approx 0.1$--$0.4$~dex). Even if we use the best-fit \sfrmodel{} calibration in the form of \cref{eq:sfr-calib-lit} ($\cha{} =-41.45$), $\sfrbias{}$ still reaches $\approx 0.3$~dex at $\tage{} \approx 2$--$3$~Myr. This bias can be explained by the fact that the existing \sfrmodel{} calibrations assume constant star formation over a long timescale \citep[typically $t=100$~Myr or $t=1$~Gyr; e.g.,][]{Hao2011}. This assumption breaks down in the simulation, where galaxies exhibit very bursty and stochastic SFHs \citep{Katz2023}.

We also examine the relationship between \sfrbias{} and stellar mass and find a negative correlation ($r=-0.44$, $p<\pcrit{}$), likely driven by the positive scaling of stellar mass with metallicity (see \citealp{Katz2023} for the discussion of the MZR in \sphinx{}). Finally, we check the dependence of \sfrbias{} on the escape fraction of ionizing radiation ($f_\mathrm{esc}$) and find a nearly flat trend ($r=-0.21$, $p=0.0009$). This result is expected because the strength of nebular emission lines is inversely proportional to the escape fraction of ionizing photons, $\propto(1-f_\mathrm{esc})$, and \sphinx{} galaxies have low $f_\mathrm{esc}$ ($f_\mathrm{esc} \lesssim 1$\%) on average. This agrees with high-redshift observations, where $f_\mathrm{esc}$ is also estimated to be low \citep[$<10$\%; e.g.,][]{Mascia24}.

\subsection{Simulation-informed \sfrmodel{} calibrations}

\begin{figure}
    \centering
    \includegraphics[width=\linewidth]{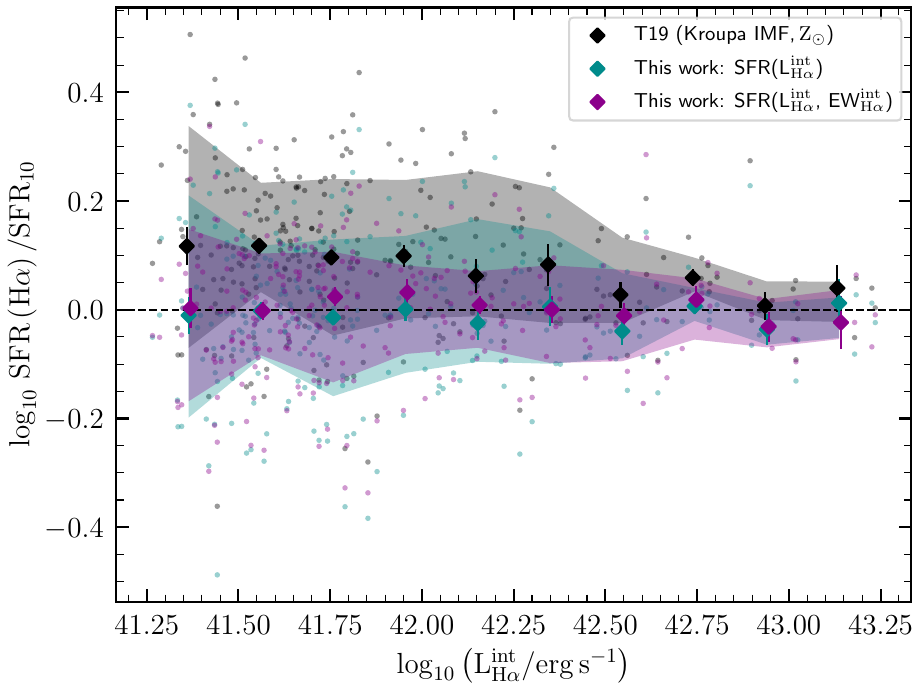}
    \caption{Comparison of the \sfrmodel{} calibrations as a function of the intrinsic \ha{} luminosity: T19 calibration (gray) and the new \sfrmodel{} calibrations given in \cref{eq:sfr-calib-lha-corr} (cyan) and \cref{eq:sfr-calib-ewha-corr} (magenta). The vertical axis represents the SFR bias factor ($\sfrbias{} \equiv \mathrm{SFR}\left(\mathrm{H}\alpha\right)/\mathrm{SFR}_{10}$; closer to $\log \sfrbias{} = 0$ is better). The diamonds show the running median, with the error bars indicating the standard error on the median. The shaded regions indicate the 68\% ($1\sigma$) confidence interval.}
    \label{fig:model-int-dsfr}
\end{figure}

Using the insights gained from our analysis of the \sfrvsha{} relation in \sphinx{}, we discuss possible improvements to the existing \sfrmodel{} calibrations that take the form of \cref{eq:sfr-calib-lit}. For example, stellar metallicity could be a useful parameter to include in the calibration to reduce the error in the predicted SFR (see \cref{fig:dsfr-vs-star-prop}). However, measuring stellar metallicity is difficult in practice, and we therefore only use quantities that can be measured robustly from galaxy spectra. Specifically, we run an ordinary least-squares (OLS) model similar to \cref{eq:sfr-calib-lit}, but allowing the \loglumhaint{} coefficient to vary. We use the randomly selected $80\%$ of the sample to train the model and the remaining $20\%$ to validate its performance. The resulting \sfrmodel{} calibration can be written as follows:
\begin{equation}
    \label{eq:sfr-calib-lha-corr}
    \begin{split}        
    \logsfrfrac{} =& \loglumhafracint{} - 41.45 \\
                   & + 0.06 \left( \loglumhafracint{} - 41.90 \right),
    \end{split}
\end{equation}
where the last term is the effective correction to \cref{eq:sfr-calib-lit}. The root mean squared error (RMSE) calculated on the training and test datasets is identical ($\mathrm{RMSE}=0.13$~dex), indicating that the model performs well on unseen data. Compared to the T19 calibration ($Z_*=Z_\odot$) which yields $\mathrm{RMSE}=0.17$~dex calculated on the full dataset, the RMSE is reduced by $\approx 0.04$~dex (see also \cref{fig:model-int-dsfr}).

The \sfrmodel{} calibration given in \cref{eq:sfr-calib-lha-corr} corrects the SFRs by a factor that depends on \lumha{} alone, which means that \sfrvar{} is neglected even when using the updated calibration. To further improve on this, we add a correction to \cref{eq:sfr-calib-lha-corr} that depends on the \ha{} equivalent width (\ewha{}). This choice is motivated by the fact that \ewha{} traces stellar metallicity and age, which both drive \sfrvar{} (\cref{fig:dsfr-vs-star-prop}). In addition, \ewha{} is only weakly sensitive to dust attenuation, which implies that dust corrections affect the results less than when other observables are used (e.g., $M_{\mathrm{UV}}$). We therefore run an OLS model that includes \ewha{}, using the same training sample as in \cref{eq:sfr-calib-lha-corr}. The resulting \sfrmodel{} calibration can be written as follows:
\begin{equation}
    \label{eq:sfr-calib-ewha-corr}
    \begin{split}        
    \logsfrfrac{} =&  \loglumhafracint{} - 41.45 \\
                   & -0.01 \left( \loglumhafracint{} - 41.90 \right) \\
                   & - 0.26 \left( \logewhafracint{} - 2.67 \right),
    \end{split}
\end{equation}
where the last two terms are the effective corrections to \cref{eq:sfr-calib-lit}. The new calibration yields $\mathrm{RMSE}=0.11$~dex on the training and test datasets, which is $\approx 0.02$~dex lower than the RMSE calculated using \cref{eq:sfr-calib-lha-corr}, or $\approx 0.06$~dex lower than the RMSE calculated using the T19 calibration ($Z_*=Z_\odot$). \Cref{fig:model-int-dsfr} compares these calibrations as a function of \lumhaint{}. In particular, this plot shows that the median \sfrbias{} in the \lumhaint{} bins is within $\pm 0.05$~dex across the entire range of luminosities ($41.25 < \loglumhaint{} < 43.25$) when using \cref{eq:sfr-calib-lha-corr} or \cref{eq:sfr-calib-ewha-corr}. This is an improvement over the T19 calibration, which overpredicts the SFRs by $\sfrbias{} \gtrsim 0.1$~dex in galaxies with low to medium luminosities ($41.25 < \loglumhaint{} \lesssim 42.5$). The standard deviation of \sfrbias{} in the \lumhaint{} bins (hereafter \sfrbiasvar{}) is nearly identical between the T19 calibration and the \sfrmodel{} calibration given in \cref{eq:sfr-calib-lha-corr} ($\sfrbiasvar{} \approx 0.11$~dex), but it decreases by $\approx 0.03$~dex ($\sfrbiasvar{} \approx 0.08$~dex) for \cref{eq:sfr-calib-ewha-corr}, especially at medium luminosities ($41.8 \lesssim \loglumhaint{} \lesssim 42.5$).

We also run a similar model in which we replace \ewha{} with \oiiihb{}, which is a proxy for the gas-phase (and, to a weaker extent, stellar) metallicity. The resulting calibration shows the same performance as \cref{eq:sfr-calib-ewha-corr} ($\mathrm{RMSE}=0.11$~dex). Given the broader wavelength coverage required to measure \oiiihb{} compared to \ewha{}, we find \cref{eq:sfr-calib-ewha-corr} to be more practical when applied to observations and hence use it as our best-performance calibration for the remainder of this paper.

\section{Discussion}
\label{sec:discussion}

\subsection{Implications}

The \ha{} line has long been known as a reliable indicator of the SFR as it traces the ionizing photon luminosity from massive ($M> 10\mathrm{M}_\odot$), short-lived ($t \lesssim 10$~Myr) stars \citep[e.g.,][]{Kennicutt1998}. Over the past decades, \ha{} has been widely employed in numerous spectroscopic \citep[e.g.,][]{Shim2009,Gunawardhana2013,Nagaraj2023} and narrow-band imaging \citep[e.g.,][]{Ly2007,Hayes2010,Sobral13} surveys to measure the cosmic SFH up to $z \sim 3$. With the recent advent of sensitive near-IR spectroscopy at $\lambda \sim 3$--$5$~$\mu$m, these measurements have been extended to $z\approx 3$--$6$, in particular thanks to the \jwst{}/NIRCam slitless spectroscopic observations of large samples of \ha{}-emitters at the same redshifts \citep[e.g.,][]{Covelo-Paz25,Fu25,Lin25}. 

As \ha{} observations become more accessible at high redshift, it is crucial to revisit the classical \sfrmodel{} calibrations \citep[e.g.,][]{Kennicutt1998} that are commonly used at $z \lesssim 3$. Based on a realistic galaxy simulation, we have shown that these calibrations bias the predicted SFRs under the physical conditions that are characteristic of high-redshift galaxies observed with \jwst{} (i.e., a low metallicity and bursty star formation), potentially also affecting population statistics. To quantify this effect, we study the effect of the metallicity- and age-dependent SFR bias (see \cref{fig:dsfr-vs-star-prop}) on the \sfrd{} measurements. Specifically, we convert the dust-corrected \ha{} luminosity function at $z \sim 6.3$ presented in Schechter form by \citet{Fu25} into the star formation rate function (SFRF). We perform this conversion by expressing the \ha{} luminosity via SFR using two \sfrvsha{} calibrations: the T19 calibration ($Z_*=Z_\odot$), and our new \sfrmodel{} calibration given in \cref{eq:sfr-calib-lha-corr}, which, by design, takes the dependence of the \sfrvsha{} relation on metallicity into account. Following previous studies, we then integrate both SFRFs down to $\mathrm{SFR}=0.24$~$M_\odot$~yr$^{-1}$ which corresponds to $M_{\mathrm{UV}}=-17$ \citep{Bouwens15}. The resulting \sfrd{} decreases from $\approx 0.017$~$M_\odot$~$\mathrm{yr}^{-1}$~$\mathrm{Mpc}^{-3}$ to $\approx 0.015$~$M_\odot$~$\mathrm{yr}^{-1}$~$\mathrm{Mpc}^{-3}$, or by $\approx 12$\%, when using the new calibration. This suggests that the effect of the evolving \sfrvsha{} relation on \sfrd{} is moderate, likely because galaxies in which the SFR bias is highest contribute little to \sfrd{}. For example, galaxies with $\mathrm{SFR}<10$~$M_\odot$~$\mathrm{yr}^{-1}$ contribute less than $30$~\% to the total \sfrd{}. If we integrate the SFRF only up to $\mathrm{SFR} = 10$~$M_\odot$~$\mathrm{yr}^{-1}$, the change in \sfrd{} would be more pronounced: \sfrd{} would decrease by $\approx 29$\%.

\begin{figure}
    \centering
    \includegraphics[width=\linewidth]{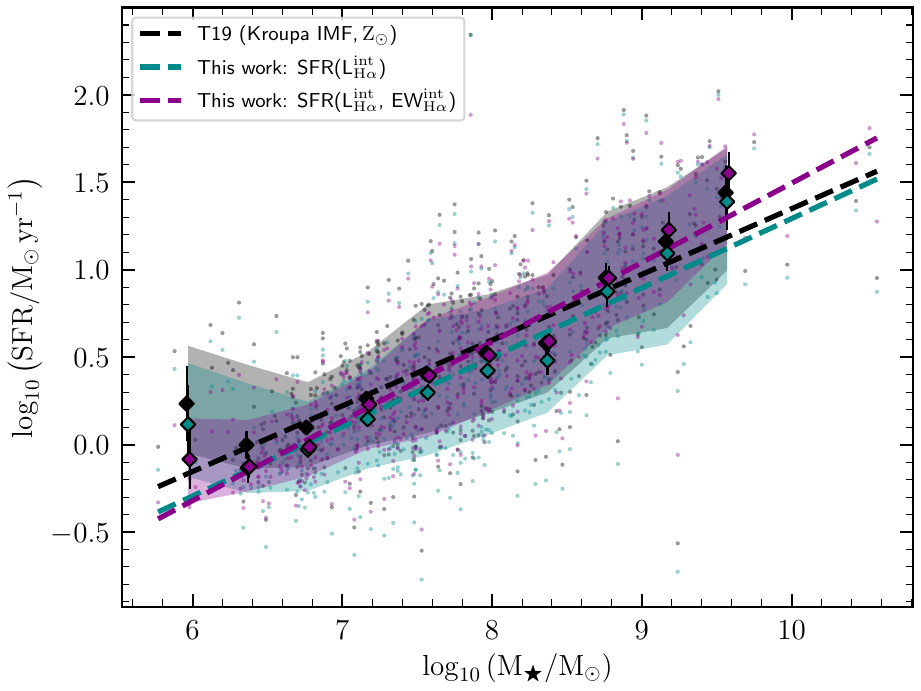}
    \caption{Star formation main sequence (\sfms{}) in the \jwst{} ALT survey \citep{Naidu24,DiCesare2025} with SFRs calculated using the T19 calibration (gray) and the new \sfrmodel{} calibrations given in \cref{eq:sfr-calib-lha-corr} (cyan) and \cref{eq:sfr-calib-ewha-corr} (magenta). The diamonds show the running median, with the error bars indicating the standard error on the median. The shaded regions indicate the 68\% ($1\sigma$) confidence interval. The dashed lines show the linear regression fits to the \sfms{} individually for each calibration.}
    \label{fig:sfms}
\end{figure}

The SFR bias also affects the correlations between the SFR and other galaxy properties, such as stellar mass. In \cref{fig:sfms}, we show the \sfrvsmstar{} relation, also known as the star formation main sequence \citep[\sfms{};][]{Noeske07}, in the \jwst{} All the Little Things (ALT) survey \citep{Naidu24,DiCesare2025}. We apply different \sfrmodel{} calibrations to the dust-corrected \ha{} luminosities and calculate the slope of the \sfms{}. The slope increases by $\Delta \partial \log \mathrm{SFR} / \partial \log \mathrm{M_\star} = 0.08 \pm 0.02$ ($20\pm6$\%) when we use \cref{eq:sfr-calib-ewha-corr} instead of the T19 calibration to calculate the SFRs. \Cref{eq:sfr-calib-lha-corr} yields similar results, although they are less significant ($\Delta \partial \log \mathrm{SFR} / \partial \log \mathrm{M_\star} = 0.02 \pm 0.03$). In addition, we find that the scatter in the \sfms{} decreases by $\Delta \sfrvar{} \approx 0.04$~dex ($\approx14$\%) when using \cref{eq:sfr-calib-ewha-corr}, most noticeably at the low-mass end ($\Delta \sfrvar{} \approx 0.06$~dex at $\log_{10} \left( M_\star/M_\odot \right) < 7$). In contrast, \cref{eq:sfr-calib-lha-corr} produces almost the same scatter as the T19 calibration ($\Delta \sfrvar{} \lesssim 0.01$~dex). We note that the \sfms{} in \cref{fig:sfms} is affected by selection effects: some low-mass galaxies are not detected in observations due to their faintness, resulting in a shallow \sfms{} slope. Temporary cessation or strong suppression of star formation in galaxies during so-called mini-quenching events \citep[e.g.,][]{Gelli2023,Looser2025} can further amplify this bias. However, a comprehensive discussion of caveats associated with the \sfms{} slope measurements is beyond the scope of this paper (this discussion can be found in \citealp{DiCesare2025}).

\subsection{Dust attenuation}
In \cref{sec:results}, we ignored the effects of dust attenuation, assuming that the H$\alpha$ line measurements (i.e., \lumha{} and \ewha{}) can be corrected for these effects without errors. This assumption is also usually made for the \sfrmodel{} calibrations presented in the literature. However, dust attenuation is known to be an important source of systematic error in the \ha{}-derived SFRs \citep[e.g.,][]{Hopkins2001}. This effect likely persists even at $z \gtrsim 4$ as the number of dust in \ha{}-emitters at these redshifts is found to be non-negligible (e.g., \citealp{Covelo-Paz25} reported the median $A_{\ha{}}\approx 0.5$~mag for a sample of $\approx 1000$ \ha{}-emitters at $4 \lesssim z \lesssim 6$).

\begin{figure}
    \centering
    \includegraphics[width=\linewidth]{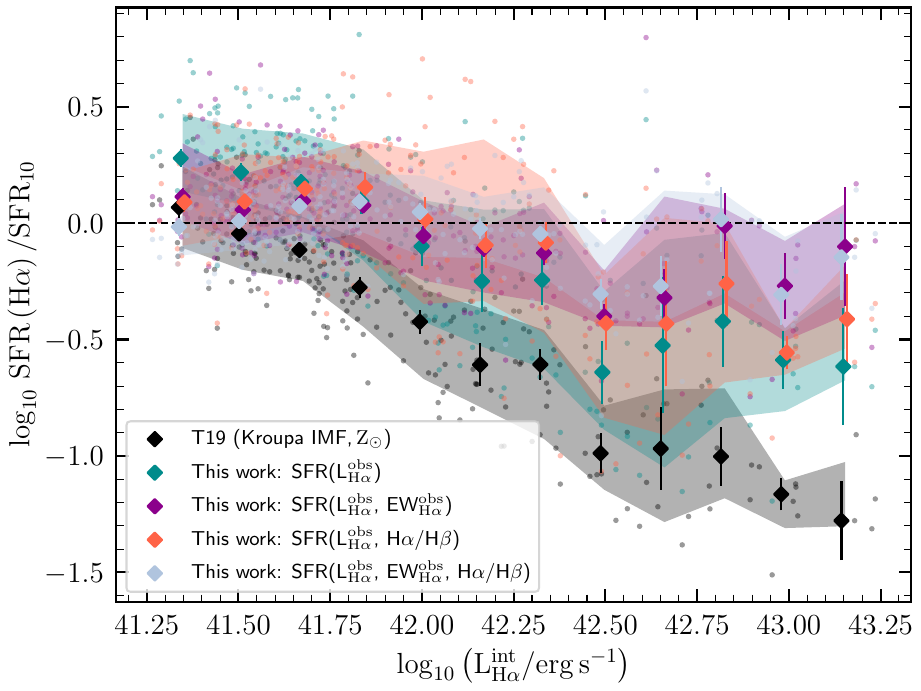}
    \caption{Same as \cref{fig:model-int-dsfr}, but for the observed (i.e., dust-attenuated) \ha{}. In addition to \lumha{} and \ewha{}, two \sfrmodel{} calibrations use the Balmer decrement (\bd{}) to predict the SFR (orange and light blue). The equations for the new calibrations are given in \cref{sec:sfr-calib-dust}.}
    \label{fig:model-obs-dsfr}
\end{figure}

Here, we test the possibility of deriving SFRs directly from the observed \ha{} emission without prior dust corrections. We use the dust-attenuated \ha{} luminosity (\lumhaobs{}) and equivalent width (\ewhaobs{}) from \sphinx{}, calculated as described in \cref{sec:methods}, and fit the same linear relations between the SFR, \lumha{}, and \ewha{} as in \cref{sec:results}. We present the resulting \sfrmodel{} calibrations in \cref{sec:sfr-calib-dust}. The calibration that only depends on \lumhaobs{} (\cref{eq:sfr-calib-lha-obs-corr}; \cref{fig:model-obs-dsfr}, cyan) yields $\mathrm{RMSE}=0.39$~dex, while the calibration that additionally depends on \ewhaobs{} (\cref{eq:sfr-calib-ewha-obs-corr}; \cref{fig:model-obs-dsfr}, magenta) yields $\mathrm{RMSE}=0.22$~dex. In both cases, the RMSE is significantly higher than the RMSEs calculated using the calibrations that are based on the intrinsic \ha{} properties (see \cref{sec:results}). Moreover, the SFRs are significantly underestimated at high luminosities ($\loglumhaint{} \gtrsim 42.5$), with \cref{eq:sfr-calib-lha-obs-corr} and \cref{eq:sfr-calib-ewha-obs-corr} producing the median $\sfrbias{} = -0.52$ and $\sfrbias{} = -0.27$, respectively. This can be explained by the fact that high-\lumhaint{} galaxies in \sphinx{} have very high attenuation values ($A_{\ha{}} \gtrsim 2$) on average, which we are unable to fully correct for while simultaneously trying to model the SFRs of low-\lumhaint{} galaxies. However, we caution that $A_{\ha{}}$ like this are not reported in high-redshift observations (see \cref{sec:methods}), and it might therefore still be possible in practice to make accurate SFR estimates based on the dust-attenuated \ha{}.

\begin{figure}
    \centering
    \includegraphics[width=\linewidth]{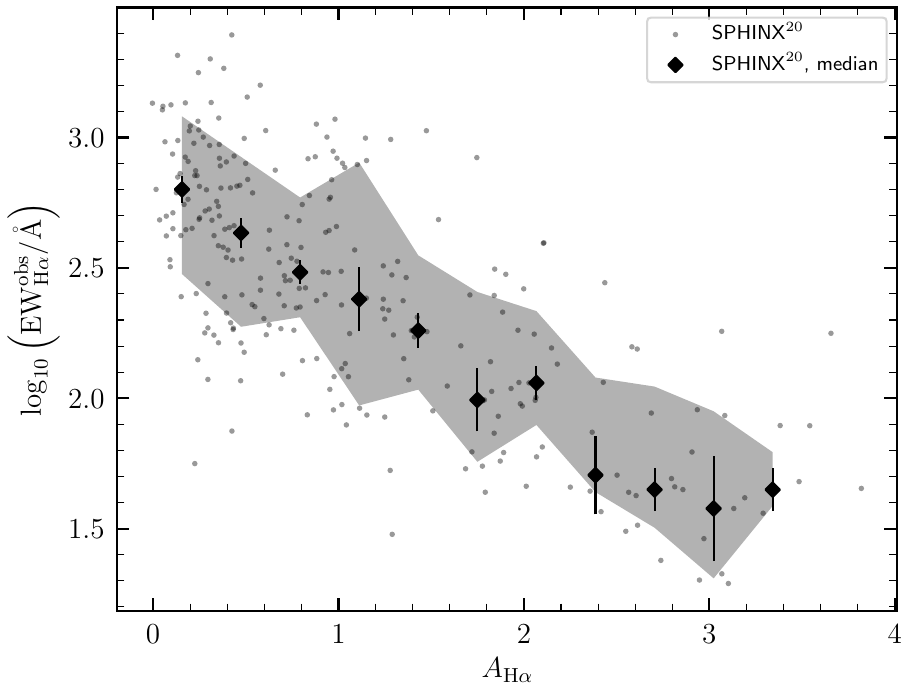}
    \caption{\ha{} equivalent width vs.~dust attenuation in \sphinx{}. The diamonds show the running median, with the error bars indicating the standard error on the median. The shaded region indicates the 68\% ($1\sigma$) confidence interval. We note that the \aha{} values from \sphinx{} might be overestimated compared to the typical \aha{} values measured in star-forming galaxies at $z \gtrsim 3$ (see \cref{sec:methods}).}
    \label{fig:ew-ha-obs-vs-a-ha}
\end{figure}

We also test whether the Balmer decrement ($\lumhaobs{}/\lumhbobs{}$; hereafter \bd{}), whose measurements are commonly used to calculate the dust corrections for \ha{}, can be a useful parameter to include in the \sfrmodel{} calibration. Interestingly, we find that the \bd{}-based calibration (\cref{eq:sfr-calib-bd-corr}; \cref{fig:model-obs-dsfr}, orange) yields $\mathrm{RMSE} = 0.29$~dex, which is $\approx0.07$~dex higher than the RMSE calculated using the \ewhaobs{}-based calibration. This indicates that \bd{} is not as powerful as \ewhaobs{} in constraining the SFRs of the \sphinx{} galaxies. This result is surprising because \bd{} is commonly regarded as a robust indicator of dust attenuation (\aha{}), while \ewhaobs{} is in principle mostly independent of it. However, we still find a negative correlation between \ewhaobs{} and \aha{} in \sphinx{} ($r=-0.71$, $p<\pcrit{}$; \cref{fig:ew-ha-obs-vs-a-ha}), possibly because the amount of dust increases with the age of the stellar population, a parameter to which \ewhaobs{} is sensitive. Further, unlike \bd{}, \ewhaobs{} correlates with stellar metallicity, allowing for an even more accurate prediction of the SFRs.

Inclusion of both \bd{} and \ewhaobs{} in the calibration (\cref{eq:sfr-calib-ewha-obs-bd-corr}; see also \cref{fig:model-obs-dsfr}, light blue) provides the best performance among the considered models ($\mathrm{RMSE} = 0.16$~dex). Observationally, however, the \bd{} measurements are hampered by slit losses and the requirement of a broader wavelength coverage. This implies that the benefits of including \bd{} in the \sfrmodel{} calibration are likely limited, with \ewhaobs{} remaining a key addition to the calibration.

\subsection{Caveats}

When using the \sfrmodel{} calibrations given in \cref{eq:sfr-calib-lha-corr} and \cref{eq:sfr-calib-ewha-corr}, it is important to take proper account of the scope of applicability of these calibrations and the assumptions made to derive them. In particular, the selection criteria described in \cref{sec:methods} imply that the new calibrations can be applied to a flux-limited sample when the following conditions are met:
\begin{itemize}
    \item $4 \lesssim z \lesssim 10$;
    \item $\lumhaobs{} \gtrsim 10^{41}$~erg~s$^{-1}$.
\end{itemize}

It is already possible to extend the calibrations to $\lumhaobs{} < 10^{41}$~erg~s$^{-1}$ with \sphinx{}, but a larger halo catalog than the currently available SPDRv1 catalog is needed to create a complete \ha{} sample down to very low luminosities. At low redshifts ($z \lesssim 4$), the SFHs become less bursty and the metallicities increase, rendering the classical calibrations more suitable. Finally, at high redshift ($z \gtrsim 10$), it remains to be established whether the new calibrations can be used, as the physical conditions of star formation at these redshifts are poorly understood.

We also stress the fact that the new calibrations assume the \bpass{} SPS model with a Kroupa-type IMF (see \cref{sec:methods}). \Cref{fig:sfr-vs-ha-int} shows that the choice of the IMF significantly affects the predicted SFRs, resulting in a systematic error of up to $ \approx 0.2$~dex (see also \citealp{Jerabkova2018}, who studied the effect of a non-universal IMF on the \sfrvsha{} relation). The choice of the SPS model, in particular, whether the model includes binary stars, results in a smaller error ($\approx 0.05$~dex), possibly because binary stars have little effect on the LyC luminosity in the first $\sim3$~Myrs of a stellar population when the bulk of ionizing photons is produced \citep[e.g.,][]{Rosdahl2018}, but this effect is still non-negligible. In the future, these systematic errors can be reduced by testing the SPS models and constraining the IMF at $z>4$, for instance, by means of deep rest-frame UV spectroscopy \citep[e.g.,][]{Steidel2016,Chisholm19}.

\section{Summary}
\label{sec:summary}

In this work, we study the relation between \ha{} emission and the SFR in the \sphinx{} cosmological simulations at $4.64 \leq z \leq 10$. Using the simulated galaxy data from the \sphinx{} Public Data Release, Version~1 \citep{Katz2023}, we select a sample of star-forming galaxies that are representative of the \ha{}-emitter population routinely observed with \jwst{} at $z > 3$. The median \sfrvsha{} relation in \sphinx{} exhibits a downturn at low luminosities (\cref{fig:sfr-vs-ha-int}; $\loglumhaint{} \sim 42.0$--$42.5$). This downturn is consistent with fainter galaxies having lower stellar metallicities ($Z_\star \sim 0.1 Z_\odot$) and thus being more efficient producers of ionizing photons than brighter, relatively metal-enriched ($Z_\star \gtrsim 0.3 Z_\odot$) galaxies. Furthermore, the \sfrvsha{} relation exhibits a significant scatter at fixed luminosity, ranging from $\sfrvar{}\approx0.04$~dex at $\loglumhaint{}=42.8$ to $\sfrvar{}\approx0.17$~dex at $\loglumhaint{}=41.3$ (i.e., the scatter is larger at lower luminosities). We demonstrate that this scatter is primarily driven by the variations in the stellar metallicities and ages (\cref{fig:dsfr-vs-star-prop}), whereas other properties (e.g., $M_{\star}$ and $f_{\mathrm{esc}}$) play a less significant role.

Because of the metallicity dependence of the \sfrvsha{} relation, the classical \sfrmodel{} calibrations (e.g., \citealp{Kennicutt1998}; \cref{eq:sfr-calib-lit}) overestimate the SFRs of faint galaxies in \sphinx{} on average by $\sfrbias{} \equiv \mathrm{SFR}(\mathrm{H}\alpha)/\mathrm{SFR}_{10} \gtrsim 0.1$~dex (\cref{fig:model-int-dsfr}, black). In \cref{eq:sfr-calib-lha-corr}, we propose a new \sfrmodel{} calibration that depends on the intrinsic \ha{} luminosity alone and reduces the error in the predicted SFRs by $\Delta \mathrm{RMSE} \approx 0.04$~dex (\cref{fig:model-int-dsfr}, cyan). To address the scatter in the \sfrvsha{} relation, we propose another \sfrmodel{} calibration in \cref{eq:sfr-calib-ewha-corr}. This calibration additionally depends on the \ha{} equivalent width, a parameter that is sensitive to the stellar metallicity and age (\cref{fig:dsfr-vs-star-prop}),  and further improves the SFR estimate, reducing the error by a total of $\Delta \mathrm{RMSE} \approx 0.06$~dex (\cref{fig:model-int-dsfr}, magenta). The new \sfrmodel{} calibrations affect the galaxy population statistics at high redshift ($z \sim 6$): the inferred $\rho_{\mathrm{SFR}}$ decreases by $\approx 12$\%, and the slope of the \sfms{} increases by $\Delta \partial \log \mathrm{SFR} / \partial \log \mathrm{M_\star} = 0.08 \pm 0.02$ (\cref{fig:sfms}).

\begin{acknowledgements}

We thank the anonymous referee for the insightful comments that helped improve the manuscript.

We also thank Thibault Garel, Pascal Oesch, Irene Shivaei, Charlotte Simmonds, Andrew Hopkins, Daniel Schaerer, and Rashmi Gottumukkala for useful comments and productive discussions.

We gratefully acknowledge support from the CBPsmn (PSMN, Pôle Scientifique de Modélisation Numérique) of the ENS de Lyon for the computing resources.

Funded by the European Union (ERC, AGENTS, 101076224). Views and opinions expressed are however those of the author(s) only and do not necessarily reflect those of the European Union or the European Research Council. Neither the European Union nor the granting authority can be held responsible for them. 

This work made extensive use of several open-source software packages, and we gratefully acknowledge the efforts of their authors: \textsc{numpy} \citep{Numpy2020}, \textsc{astropy} \citep{Astropy2022}, \textsc{matplotlib} \citep{Matplotlib2007}, \textsc{ipython} \citep{IPython2007}, and \textsc{scikit-learn} \citep{scikit-learn}.

\end{acknowledgements}

\bibliographystyle{aa}
\bibliography{aa57114-25}

@ARTICLE{Gelli2023,
       author = {{Gelli}, Viola and {Salvadori}, Stefania and {Ferrara}, Andrea and {Pallottini}, Andrea and {Carniani}, Stefano},
        title = "{Quiescent Low-mass Galaxies Observed by JWST in the Epoch of Reionization}",
      journal = {\apjl},
         year = 2023,
        month = sep,
       volume = {954},
       number = {1},
          eid = {L11},
        pages = {L11},
          doi = {10.3847/2041-8213/acee80},
archivePrefix = {arXiv},
       eprint = {2303.13574},
 primaryClass = {astro-ph.GA},
       adsurl = {https://ui.adsabs.harvard.edu/abs/2023ApJ...954L..11G}
}

@ARTICLE{DiCesare2025,
       author = {{Di Cesare}, Claudia and {Matthee}, Jorryt and {Naidu}, Rohan P. and {Torralba}, Alberto and {Kotiwale}, Gauri and {Kramarenko}, Ivan G. and {Blazoit}, Jeremy and {Rosdahl}, Joakim and {Leja}, Joel and {Iani}, Edoardo and {Adamo}, Angela and {Covelo-Paz}, Alba and {Furtak}, Lukas J. and {Heintz}, Kasper E. and {Mascia}, Sara and {Navarrete}, Benjam{\'\i}n and {Oesch}, Pascal A. and {Romano}, Michael and {Shivaei}, Irene and {Tacchella}, Sandro},
        title = "{The slope and scatter of the star forming main sequence at z\raisebox{-0.5ex}\textasciitilde5 : reconciling observations with simulations}",
      journal = {arXiv e-prints},
         year = 2025,
        month = oct,
          eid = {arXiv:2510.19044},
        pages = {arXiv:2510.19044},
          doi = {10.48550/arXiv.2510.19044},
archivePrefix = {arXiv},
       eprint = {2510.19044},
 primaryClass = {astro-ph.GA},
       adsurl = {https://ui.adsabs.harvard.edu/abs/2025arXiv251019044D}
}

@ARTICLE{Jerabkova2018,
       author = {{Je{\v{r}}{\'a}bkov{\'a}}, T. and {Zonoozi}, A. Hasani and {Kroupa}, P. and {Beccari}, G. and {Yan}, Z. and {Vazdekis}, A. and {Zhang}, Z.-Y.},
        title = "{Impact of metallicity and star formation rate on the time-dependent, galaxy-wide stellar initial mass function}",
      journal = {\aap},
         year = 2018,
        month = nov,
       volume = {620},
          eid = {A39},
        pages = {A39},
          doi = {10.1051/0004-6361/201833055},
archivePrefix = {arXiv},
       eprint = {1809.04603},
 primaryClass = {astro-ph.GA},
       adsurl = {https://ui.adsabs.harvard.edu/abs/2018A&A...620A..39J}
}

@ARTICLE{Markov2025,
       author = {{Markov}, Vladan and {Gallerani}, Simona and {Ferrara}, Andrea and {Pallottini}, Andrea and {Parlanti}, Eleonora and {Mascia}, Fabio Di and {Sommovigo}, Laura and {Kohandel}, Mahsa},
        title = "{The evolution of dust attenuation in z {\ensuremath{\approx}} 2-12 galaxies observed by JWST}",
      journal = {Nature Astronomy},
         year = 2025,
        month = mar,
       volume = {9},
        pages = {458-468},
          doi = {10.1038/s41550-024-02426-1},
archivePrefix = {arXiv},
       eprint = {2402.05996},
 primaryClass = {astro-ph.GA},
       adsurl = {https://ui.adsabs.harvard.edu/abs/2025NatAs...9..458M}
}

@ARTICLE{Laursen2009,
       author = {{Laursen}, Peter and {Sommer-Larsen}, Jesper and {Andersen}, Anja C.},
        title = "{Ly{\ensuremath{\alpha}} Radiative Transfer with Dust: Escape Fractions from Simulated High-Redshift Galaxies}",
      journal = {\apj},
         year = 2009,
        month = oct,
       volume = {704},
       number = {2},
        pages = {1640-1656},
          doi = {10.1088/0004-637X/704/2/1640},
archivePrefix = {arXiv},
       eprint = {0907.2698},
 primaryClass = {astro-ph.CO},
       adsurl = {https://ui.adsabs.harvard.edu/abs/2009ApJ...704.1640L}
}

@ARTICLE{Bezanson2024,
       author = {{Bezanson}, Rachel and {Labbe}, Ivo and {Whitaker}, Katherine E. and {Leja}, Joel and {Price}, Sedona H. and {Franx}, Marijn and {Brammer}, Gabriel and {Marchesini}, Danilo and {Zitrin}, Adi and {Wang}, Bingjie and {Weaver}, John R. and {Furtak}, Lukas J. and {Atek}, Hakim and {Coe}, Dan and {Cutler}, Sam E. and {Dayal}, Pratika and {van Dokkum}, Pieter and {Feldmann}, Robert and {F{\"o}rster Schreiber}, Natascha M. and {Fujimoto}, Seiji and {Geha}, Marla and {Glazebrook}, Karl and {de Graaff}, Anna and {Greene}, Jenny E. and {Juneau}, St{\'e}phanie and {Kassin}, Susan and {Kriek}, Mariska and {Khullar}, Gourav and {Maseda}, Michael and {Mowla}, Lamiya A. and {Muzzin}, Adam and {Nanayakkara}, Themiya and {Nelson}, Erica J. and {Oesch}, Pascal A. and {Pacifici}, Camilla and {Pan}, Richard and {Papovich}, Casey and {Setton}, David J. and {Shapley}, Alice E. and {Smit}, Renske and {Stefanon}, Mauro and {Taylor}, Edward N. and {Williams}, Christina C.},
        title = "{The JWST UNCOVER Treasury Survey: Ultradeep NIRSpec and NIRCam Observations before the Epoch of Reionization}",
      journal = {\apj},
         year = 2024,
        month = oct,
       volume = {974},
       number = {1},
          eid = {92},
        pages = {92},
          doi = {10.3847/1538-4357/ad66cf},
archivePrefix = {arXiv},
       eprint = {2212.04026},
 primaryClass = {astro-ph.GA},
 primaryClass = {astro-ph.GA},
       adsurl = {https://ui.adsabs.harvard.edu/abs/2024ApJ...974...92B}
}

@ARTICLE{Shivaei2025,
       author = {{Shivaei}, Irene and {Naidu}, Rohan P. and {Rodr{\'\i}guez Montero}, Francisco and {Matsumoto}, Kosei and {Leja}, Joel and {Matthee}, Jorryt and {Johnson}, Benjamin D. and {Oesch}, Pascal A. and {Chevallard}, Jacopo and {Adamo}, Angela and {Bodansky}, Sarah and {Bunker}, Andrew J. and {Covelo Paz}, Alba and {Di Cesare}, Claudia and {Egami}, Eiichi and {Furtak}, Lukas J. and {Heintz}, Kasper E. and {Kramarenko}, Ivan and {Meyer}, Romain A. and {Reddy}, Naveen A. and {Rinaldi}, Pierluigi and {Tacchella}, Sandro and {Torralba}, Alberto and {Witstok}, Joris and {Wozniak}, Michael A. and {Xiao}, Mengyuan},
        title = "{The Diversity and Evolution of Dust Attenuation Curves from Redshift z \raisebox{-0.5ex}\textasciitilde 1 to 9}",
      journal = {arXiv e-prints},
         year = 2025,
        month = sep,
          eid = {arXiv:2509.01795},
        pages = {arXiv:2509.01795},
          doi = {10.48550/arXiv.2509.01795},
archivePrefix = {arXiv},
       eprint = {2509.01795},
 primaryClass = {astro-ph.GA},
       adsurl = {https://ui.adsabs.harvard.edu/abs/2025arXiv250901795S}
}

@ARTICLE{Fisher2025,
       author = {{Fisher}, R. and {Bowler}, R.~A.~A. and {Stefanon}, M. and {Rowland}, L.~E. and {Algera}, H.~S.~B. and {Aravena}, M. and {Bouwens}, R. and {Dayal}, P. and {Ferrara}, A. and {Fudamoto}, Y. and {Gulis}, C. and {Hodge}, J.~A. and {Inami}, H. and {Ormerod}, K. and {Pallottini}, A. and {Phillips}, S.~G. and {Sartorio}, N.~S. and {Schouws}, S. and {Smit}, R. and {Sommovigo}, L. and {Stark}, D.~P. and {van der Werf}, P.~P.},
        title = "{REBELS-IFU: dust attenuation curves of 12 massive galaxies at z ≃ 7}",
      journal = {\mnras},
         year = 2025,
        month = may,
       volume = {539},
       number = {1},
        pages = {109-126},
          doi = {10.1093/mnras/staf485},
archivePrefix = {arXiv},
       eprint = {2501.10541},
 primaryClass = {astro-ph.GA},
       adsurl = {https://ui.adsabs.harvard.edu/abs/2025MNRAS.539..109F}
}

@ARTICLE{Wilkins2019,
       author = {{Wilkins}, Stephen M. and {Lovell}, Christopher C. and {Stanway}, Elizabeth R.},
        title = "{Recalibrating the cosmic star formation history}",
      journal = {\mnras},
         year = 2019,
        month = dec,
       volume = {490},
       number = {4},
        pages = {5359-5365},
          doi = {10.1093/mnras/stz2894},
archivePrefix = {arXiv},
       eprint = {1910.05220},
 primaryClass = {astro-ph.GA},
       adsurl = {https://ui.adsabs.harvard.edu/abs/2019MNRAS.490.5359W}
}

@ARTICLE{Teyssier2002,
       author = {{Teyssier}, R.},
        title = "{Cosmological hydrodynamics with adaptive mesh refinement. A new high resolution code called RAMSES}",
      journal = {\aap},
         year = 2002,
        month = apr,
       volume = {385},
        pages = {337-364},
          doi = {10.1051/0004-6361:20011817},
archivePrefix = {arXiv},
       eprint = {astro-ph/0111367},
 primaryClass = {astro-ph},
       adsurl = {https://ui.adsabs.harvard.edu/abs/2002A&A...385..337T}
}

@ARTICLE{DeLooze2014,
       author = {{De Looze}, Ilse and {Cormier}, Diane and {Lebouteiller}, Vianney and {Madden}, Suzanne and {Baes}, Maarten and {Bendo}, George J. and {Boquien}, M{\'e}d{\'e}ric and {Boselli}, Alessandro and {Clements}, David L. and {Cortese}, Luca and {Cooray}, Asantha and {Galametz}, Maud and {Galliano}, Fr{\'e}d{\'e}ric and {Graci{\'a}-Carpio}, Javier and {Isaak}, Kate and {Karczewski}, Oskar {\L}. and {Parkin}, Tara J. and {Pellegrini}, Eric W. and {R{\'e}my-Ruyer}, Aur{\'e}lie and {Spinoglio}, Luigi and {Smith}, Matthew W.~L. and {Sturm}, Eckhard},
        title = "{The applicability of far-infrared fine-structure lines as star formation rate tracers over wide ranges of metallicities and galaxy types}",
      journal = {\aap},
         year = 2014,
        month = aug,
       volume = {568},
          eid = {A62},
        pages = {A62},
          doi = {10.1051/0004-6361/201322489},
archivePrefix = {arXiv},
       eprint = {1402.4075},
 primaryClass = {astro-ph.GA},
       adsurl = {https://ui.adsabs.harvard.edu/abs/2014A&A...568A..62D}
}

@ARTICLE{Carvajal-Bohorquez2025,
       author = {{Carvajal-Bohorquez}, C. and {Ciesla}, L. and {Laporte}, N. and {Boquien}, M. and {Buat}, V. and {Ilbert}, O. and {Aufort}, G. and {Shuntov}, M. and {Witten}, C. and {Oesch}, P.~A. and {Covelo-Paz}, A.},
        title = "{Stochastic star formation activity of galaxies within the first billion years probed by JWST}",
      journal = {\aap},
         year = 2025,
        month = dec,
       volume = {704},
          eid = {A290},
        pages = {A290},
          doi = {10.1051/0004-6361/202556471},
archivePrefix = {arXiv},
       eprint = {2507.13160},
 primaryClass = {astro-ph.GA},
       adsurl = {https://ui.adsabs.harvard.edu/abs/2025A&A...704A.290C}
}

@ARTICLE{Meyer2024,
       author = {{Meyer}, R.~A. and {Oesch}, P.~A. and {Giovinazzo}, E. and {Weibel}, A. and {Brammer}, G. and {Matthee}, J. and {Naidu}, R.~P. and {Bouwens}, R.~J. and {Chisholm}, J. and {Covelo-Paz}, A. and {Fudamoto}, Y. and {Maseda}, M. and {Nelson}, E. and {Shivaei}, I. and {Xiao}, M. and {Herard-Demanche}, T. and {Illingworth}, G.~D. and {Kerutt}, J. and {Kramarenko}, I. and {Labbe}, I. and {Leonova}, E. and {Magee}, D. and {Matharu}, J. and {Prieto Lyon}, G. and {Reddy}, N. and {Schaerer}, D. and {Shapley}, A. and {Stefanon}, M. and {Wozniak}, M.~A. and {Wuyts}, S.},
        title = "{JWST FRESCO: a comprehensive census of H {\ensuremath{\beta}} + [O III] emitters at 6.8 < z < 9.0 in the GOODS fields}",
      journal = {\mnras},
         year = 2024,
        month = nov,
       volume = {535},
       number = {1},
        pages = {1067-1094},
          doi = {10.1093/mnras/stae2353},
archivePrefix = {arXiv},
       eprint = {2405.05111},
 primaryClass = {astro-ph.GA},
       adsurl = {https://ui.adsabs.harvard.edu/abs/2024MNRAS.535.1067M}
}

@ARTICLE{Weibel2024,
       author = {{Weibel}, Andrea and {Oesch}, Pascal A. and {Barrufet}, Laia and {Gottumukkala}, Rashmi and {Ellis}, Richard S. and {Santini}, Paola and {Weaver}, John R. and {Allen}, Natalie and {Bouwens}, Rychard and {Bowler}, Rebecca A.~A. and {Brammer}, Gabe and {Carnall}, Adam C. and {Cullen}, Fergus and {Dayal}, Pratika and {Dickinson}, Mark and {Donnan}, Callum T. and {Dunlop}, James S. and {Giavalisco}, Mauro and {Grogin}, Norman A. and {Illingworth}, Garth D. and {Koekemoer}, Anton M. and {Labbe}, Ivo and {Marchesini}, Danilo and {McLeod}, Derek J. and {McLure}, Ross J. and {Naidu}, Rohan P. and {P{\'e}rez-Gonz{\'a}lez}, Pablo G. and {Shuntov}, Marko and {Stefanon}, Mauro and {Toft}, Sune and {Xiao}, Mengyuan},
        title = "{Galaxy build-up in the first 1.5 Gyr of cosmic history: insights from the stellar mass function at z   4-9 from JWST NIRCam observations}",
      journal = {\mnras},
         year = 2024,
        month = sep,
       volume = {533},
       number = {2},
        pages = {1808-1838},
          doi = {10.1093/mnras/stae1891},
archivePrefix = {arXiv},
       eprint = {2403.08872},
 primaryClass = {astro-ph.GA},
       adsurl = {https://ui.adsabs.harvard.edu/abs/2024MNRAS.533.1808W}
}

@ARTICLE{Rosdahl2013,
       author = {{Rosdahl}, J. and {Blaizot}, J. and {Aubert}, D. and {Stranex}, T. and {Teyssier}, R.},
        title = "{RAMSES-RT: radiation hydrodynamics in the cosmological context}",
      journal = {\mnras},
         year = 2013,
        month = dec,
       volume = {436},
       number = {3},
        pages = {2188-2231},
          doi = {10.1093/mnras/stt1722},
archivePrefix = {arXiv},
       eprint = {1304.7126},
 primaryClass = {astro-ph.CO},
       adsurl = {https://ui.adsabs.harvard.edu/abs/2013MNRAS.436.2188R}
}

@ARTICLE{Rosdahl2022,
       author = {{Rosdahl}, Joakim and {Blaizot}, J{\'e}r{\'e}my and {Katz}, Harley and {Kimm}, Taysun and {Garel}, Thibault and {Haehnelt}, Martin and {Keating}, Laura C. and {Martin-Alvarez}, Sergio and {Michel-Dansac}, L{\'e}o and {Ocvirk}, Pierre},
        title = "{LyC escape from SPHINX galaxies in the Epoch of Reionization}",
      journal = {\mnras},
         year = 2022,
        month = sep,
       volume = {515},
       number = {2},
        pages = {2386-2414},
          doi = {10.1093/mnras/stac1942},
archivePrefix = {arXiv},
       eprint = {2207.03232},
 primaryClass = {astro-ph.GA},
       adsurl = {https://ui.adsabs.harvard.edu/abs/2022MNRAS.515.2386R}
}

@ARTICLE{Chisholm19,
       author = {{Chisholm}, J. and {Rigby}, J.~R. and {Bayliss}, M. and {Berg}, D.~A. and {Dahle}, H. and {Gladders}, M. and {Sharon}, K.},
        title = "{Constraining the Metallicities, Ages, Star Formation Histories, and Ionizing Continua of Extragalactic Massive Star Populations}",
      journal = {\apj},
         year = 2019,
        month = sep,
       volume = {882},
       number = {2},
          eid = {182},
        pages = {182},
          doi = {10.3847/1538-4357/ab3104},
archivePrefix = {arXiv},
       eprint = {1905.04314},
 primaryClass = {astro-ph.GA},
       adsurl = {https://ui.adsabs.harvard.edu/abs/2019ApJ...882..182C}
}

@ARTICLE{Michel-Dansac2020,
       author = {{Michel-Dansac}, L. and {Blaizot}, J. and {Garel}, T. and {Verhamme}, A. and {Kimm}, T. and {Trebitsch}, M.},
        title = "{RASCAS: RAdiation SCattering in Astrophysical Simulations}",
      journal = {\aap},
         year = 2020,
        month = mar,
       volume = {635},
          eid = {A154},
        pages = {A154},
          doi = {10.1051/0004-6361/201834961},
archivePrefix = {arXiv},
       eprint = {2001.11252},
 primaryClass = {astro-ph.GA},
       adsurl = {https://ui.adsabs.harvard.edu/abs/2020A&A...635A.154M}
}

@ARTICLE{Ferland2017,
       author = {{Ferland}, G.~J. and {Chatzikos}, M. and {Guzm{\'a}n}, F. and {Lykins}, M.~L. and {van Hoof}, P.~A.~M. and {Williams}, R.~J.~R. and {Abel}, N.~P. and {Badnell}, N.~R. and {Keenan}, F.~P. and {Porter}, R.~L. and {Stancil}, P.~C.},
        title = "{The 2017 Release Cloudy}",
      journal = {\rmxaa},
         year = 2017,
        month = oct,
       volume = {53},
        pages = {385-438},
          doi = {10.48550/arXiv.1705.10877},
archivePrefix = {arXiv},
       eprint = {1705.10877},
 primaryClass = {astro-ph.GA},
       adsurl = {https://ui.adsabs.harvard.edu/abs/2017RMxAA..53..385F}
}

@ARTICLE{Hopkins2001,
       author = {{Hopkins}, A.~M. and {Connolly}, A.~J. and {Haarsma}, D.~B. and {Cram}, L.~E.},
        title = "{Toward a Resolution of the Discrepancy between Different Estimators of Star Formation Rate}",
      journal = {\aj},
         year = 2001,
        month = jul,
       volume = {122},
       number = {1},
        pages = {288-296},
          doi = {10.1086/321113},
archivePrefix = {arXiv},
       eprint = {astro-ph/0103253},
 primaryClass = {astro-ph},
       adsurl = {https://ui.adsabs.harvard.edu/abs/2001AJ....122..288H}
}

@ARTICLE{Gunawardhana2013,
       author = {{Gunawardhana}, M.~L.~P. and {Hopkins}, A.~M. and {Bland-Hawthorn}, J. and {Brough}, S. and {Sharp}, R. and {Loveday}, J. and {Taylor}, E. and {Jones}, D.~H. and {Lara-L{\'o}pez}, M.~A. and {Bauer}, A.~E. and {Colless}, M. and {Owers}, M. and {Baldry}, I.~K. and {L{\'o}pez-S{\'a}nchez}, A.~R. and {Foster}, C. and {Bamford}, S. and {Brown}, M.~J.~I. and {Driver}, S.~P. and {Drinkwater}, M.~J. and {Liske}, J. and {Meyer}, M. and {Norberg}, P. and {Robotham}, A.~S.~G. and {Ching}, J.~H.~Y. and {Cluver}, M.~E. and {Croom}, S. and {Kelvin}, L. and {Prescott}, M. and {Steele}, O. and {Thomas}, D. and {Wang}, L.},
        title = "{Galaxy And Mass Assembly: evolution of the H{\ensuremath{\alpha}} luminosity function and star formation rate density up to z < 0.35}",
      journal = {\mnras},
         year = 2013,
        month = aug,
       volume = {433},
       number = {4},
        pages = {2764-2789},
          doi = {10.1093/mnras/stt890},
archivePrefix = {arXiv},
       eprint = {1305.5308},
 primaryClass = {astro-ph.CO},
       adsurl = {https://ui.adsabs.harvard.edu/abs/2013MNRAS.433.2764G}
}

@ARTICLE{Looser2025,
       author = {{Looser}, Tobias J. and {D'Eugenio}, Francesco and {Maiolino}, Roberto and {Tacchella}, Sandro and {Curti}, Mirko and {Arribas}, Santiago and {Baker}, William M. and {Baum}, Stefi and {Bonaventura}, Nina and {Boyett}, Kristan and {Bunker}, Andrew J. and {Carniani}, Stefano and {Charlot}, Stephane and {Chevallard}, Jacopo and {Curtis-Lake}, Emma and {Lola Danhaive}, A. and {Eisenstein}, Daniel J. and {de Graaff}, Anna and {Hainline}, Kevin and {Ji}, Zhiyuan and {Johnson}, Benjamin D. and {Kumari}, Nimisha and {Nelson}, Erica and {Parlanti}, Eleonora and {Rix}, Hans-Walter and {Robertson}, Brant and {Del Pino}, Bruno Rodr{\'\i}guez and {Sandles}, Lester and {Scholtz}, Jan and {Smit}, Renske and {Stark}, Daniel P. and {{\"U}bler}, Hannah and {Williams}, Christina C. and {Willott}, Chris and {Witstok}, Joris},
        title = "{JADES: Differing assembly histories of galaxies: Observational evidence for bursty star formation histories and (mini-)quenching in the first billion years of the Universe}",
      journal = {\aap},
         year = 2025,
        month = may,
       volume = {697},
          eid = {A88},
        pages = {A88},
          doi = {10.1051/0004-6361/202347102},
archivePrefix = {arXiv},
       eprint = {2306.02470},
 primaryClass = {astro-ph.GA},
       adsurl = {https://ui.adsabs.harvard.edu/abs/2025A&A...697A..88L}
}

@ARTICLE{Ly2007,
       author = {{Ly}, Chun and {Malkan}, Matt A. and {Kashikawa}, Nobunari and {Shimasaku}, Kazuhiro and {Doi}, Mamoru and {Nagao}, Tohru and {Iye}, Masanori and {Kodama}, Tadayuki and {Morokuma}, Tomoki and {Motohara}, Kentaro},
        title = "{The Luminosity Function and Star Formation Rate between Redshifts of 0.07 and 1.47 for Narrowband Emitters in the Subaru Deep Field}",
      journal = {\apj},
         year = 2007,
        month = mar,
       volume = {657},
       number = {2},
        pages = {738-759},
          doi = {10.1086/510828},
archivePrefix = {arXiv},
       eprint = {astro-ph/0610846},
 primaryClass = {astro-ph},
       adsurl = {https://ui.adsabs.harvard.edu/abs/2007ApJ...657..738L}
}

@ARTICLE{Shim2009,
       author = {{Shim}, Hyunjin and {Colbert}, James and {Teplitz}, Harry and {Henry}, Alaina and {Malkan}, Mattew and {McCarthy}, Patrick and {Yan}, Lin},
        title = "{Global Star Formation Rate Density over 0.7 < z < 1.9}",
      journal = {\apj},
         year = 2009,
        month = may,
       volume = {696},
       number = {1},
        pages = {785-796},
          doi = {10.1088/0004-637X/696/1/785},
archivePrefix = {arXiv},
       eprint = {0902.0736},
 primaryClass = {astro-ph.CO},
       adsurl = {https://ui.adsabs.harvard.edu/abs/2009ApJ...696..785S}
}

@article{Nagaraj2023,
doi = {10.3847/1538-4357/aca7bb},
url = {https://dx.doi.org/10.3847/1538-4357/aca7bb},
year = {2023},
month = {jan},
publisher = {The American Astronomical Society},
volume = {943},
number = {1},
pages = {5},
author = {Nagaraj, Gautam and Ciardullo, Robin and Bowman, William P. and Lawson, Alex and Gronwall, Caryl},
title = {The Hα and [O iii] λ5007 Luminosity Functions of 1.2 &lt; z &lt; 1.9 Emission-line Galaxies from Hubble Space Telescope (HST) Grism Spectroscopy},
journal = {The Astrophysical Journal}
}

@ARTICLE{Hayes2010,
       author = {{Hayes}, M. and {Schaerer}, D. and {{\"O}stlin}, G.},
        title = "{The H-alpha luminosity function at redshift 2.2 . A new determination using VLT/HAWK-I}",
      journal = {\aap},
         year = 2010,
        month = jan,
       volume = {509},
          eid = {L5},
        pages = {L5},
          doi = {10.1051/0004-6361/200913217},
archivePrefix = {arXiv},
       eprint = {0912.3267},
 primaryClass = {astro-ph.CO},
       adsurl = {https://ui.adsabs.harvard.edu/abs/2010A&A...509L...5H}
}

@ARTICLE{Steidel2016,
       author = {{Steidel}, Charles C. and {Strom}, Allison L. and {Pettini}, Max and {Rudie}, Gwen C. and {Reddy}, Naveen A. and {Trainor}, Ryan F.},
        title = "{Reconciling the Stellar and Nebular Spectra of High-redshift Galaxies}",
      journal = {\apj},
         year = 2016,
        month = aug,
       volume = {826},
       number = {2},
          eid = {159},
        pages = {159},
          doi = {10.3847/0004-637X/826/2/159},
archivePrefix = {arXiv},
       eprint = {1605.07186},
 primaryClass = {astro-ph.GA},
       adsurl = {https://ui.adsabs.harvard.edu/abs/2016ApJ...826..159S}
}

@ARTICLE{Lilly13,
       author = {{Lilly}, Simon J. and {Carollo}, C. Marcella and {Pipino}, Antonio and {Renzini}, Alvio and {Peng}, Yingjie},
        title = "{Gas Regulation of Galaxies: The Evolution of the Cosmic Specific Star Formation Rate, the Metallicity-Mass-Star-formation Rate Relation, and the Stellar Content of Halos}",
      journal = {\apj},
         year = 2013,
        month = aug,
       volume = {772},
       number = {2},
          eid = {119},
        pages = {119},
          doi = {10.1088/0004-637X/772/2/119},
archivePrefix = {arXiv},
       eprint = {1303.5059},
 primaryClass = {astro-ph.CO},
       adsurl = {https://ui.adsabs.harvard.edu/abs/2013ApJ...772..119L}
}

@ARTICLE{Behroozi19,
       author = {{Behroozi}, Peter and {Wechsler}, Risa H. and {Hearin}, Andrew P. and {Conroy}, Charlie},
        title = "{UNIVERSEMACHINE: The correlation between galaxy growth and dark matter halo assembly from z = 0-10}",
      journal = {\mnras},
         year = 2019,
        month = sep,
       volume = {488},
       number = {3},
        pages = {3143-3194},
          doi = {10.1093/mnras/stz1182},
archivePrefix = {arXiv},
       eprint = {1806.07893},
 primaryClass = {astro-ph.GA},
       adsurl = {https://ui.adsabs.harvard.edu/abs/2019MNRAS.488.3143B}
}

@ARTICLE{Clarke24,
       author = {{Clarke}, Leonardo and {Shapley}, Alice E. and {Sanders}, Ryan L. and {Topping}, Michael W. and {Brammer}, Gabriel B. and {Bento}, Trinity and {Reddy}, Naveen A. and {Kehoe}, Emily},
        title = "{The Star-forming Main Sequence in JADES and CEERS at z > 1.4: Investigating the Burstiness of Star Formation}",
      journal = {\apj},
         year = 2024,
        month = dec,
       volume = {977},
       number = {1},
          eid = {133},
        pages = {133},
          doi = {10.3847/1538-4357/ad8ba4},
archivePrefix = {arXiv},
       eprint = {2406.05178},
 primaryClass = {astro-ph.GA},
       adsurl = {https://ui.adsabs.harvard.edu/abs/2024ApJ...977..133C}
}

@ARTICLE{Chruslinska24,
       author = {{Chru{\'s}li{\'n}ska}, M. and {Pakmor}, R. and {Matthee}, J. and {Matsuno}, T.},
        title = "{Trading oxygen for iron. I. The [O/Fe]-specific star formation rate relation of galaxies}",
      journal = {\aap},
         year = 2024,
        month = jun,
       volume = {686},
          eid = {A186},
        pages = {A186},
          doi = {10.1051/0004-6361/202347602},
archivePrefix = {arXiv},
       eprint = {2308.00023},
 primaryClass = {astro-ph.GA},
       adsurl = {https://ui.adsabs.harvard.edu/abs/2024A&A...686A.186C}
}

@ARTICLE{Noeske07,
       author = {{Noeske}, K.~G. and {Weiner}, B.~J. and {Faber}, S.~M. and {Papovich}, C. and {Koo}, D.~C. and {Somerville}, R.~S. and {Bundy}, K. and {Conselice}, C.~J. and {Newman}, J.~A. and {Schiminovich}, D. and {Le Floc'h}, E. and {Coil}, A.~L. and {Rieke}, G.~H. and {Lotz}, J.~M. and {Primack}, J.~R. and {Barmby}, P. and {Cooper}, M.~C. and {Davis}, M. and {Ellis}, R.~S. and {Fazio}, G.~G. and {Guhathakurta}, P. and {Huang}, J. and {Kassin}, S.~A. and {Martin}, D.~C. and {Phillips}, A.~C. and {Rich}, R.~M. and {Small}, T.~A. and {Willmer}, C.~N.~A. and {Wilson}, G.},
        title = "{Star Formation in AEGIS Field Galaxies since z=1.1: The Dominance of Gradually Declining Star Formation, and the Main Sequence of Star-forming Galaxies}",
      journal = {\apjl},
         year = 2007,
        month = may,
       volume = {660},
       number = {1},
        pages = {L43-L46},
          doi = {10.1086/517926},
archivePrefix = {arXiv},
       eprint = {astro-ph/0701924},
 primaryClass = {astro-ph},
       adsurl = {https://ui.adsabs.harvard.edu/abs/2007ApJ...660L..43N}
}

@ARTICLE{Fu25,
       author = {{Fu}, Shuqi and {Sun}, Fengwu and {Jiang}, Linhua and {Lin}, Xiaojing and {Diego}, Jose M. and {Furtak}, Lukas J. and {Jauzac}, Mathilde and {Koekemoer}, Anton M. and {Li}, Mingyu and {Oguri}, Masamune and {Patel}, Nency R. and {Willmer}, Christopher N.~A. and {Windhorst}, Rogier A. and {Zitrin}, Adi and {Bauer}, Franz E. and {Chen}, Chian-Chou and {Chen}, Wenlei and {Cheng}, Cheng and {Conselice}, Christopher J. and {Eisenstein}, Daniel J. and {Egami}, Eiichi and {Espada}, Daniel and {Fan}, Xiaohui and {Fujimoto}, Seiji and {Hsiao}, Tiger Yu-Yang and {Jin}, Xiangyu and {Kohno}, Kotaro and {Lagattuta}, David J. and {Li}, Zihao and {Liu}, Weizhe and {Miralda-Escud{\'e}}, Jordi and {Ning}, Yuanhang and {Tacchella}, Sandro and {Tee}, Wei Leong and {Umehata}, Hideki and {Wang}, Feige and {Yan}, Haojing and {Zhu}, Yongda},
        title = "{Medium-band Astrophysics with the Grism of NIRCam In Frontier Fields (MAGNIF): Spectroscopic Census of H{\ensuremath{\alpha}} Luminosity Functions and Cosmic Star Formation at z {\ensuremath{\sim}} 4.5 and 6.3}",
      journal = {\apj},
         year = 2025,
        month = jul,
       volume = {987},
       number = {2},
          eid = {186},
        pages = {186},
          doi = {10.3847/1538-4357/adddb1},
archivePrefix = {arXiv},
       eprint = {2503.03829},
 primaryClass = {astro-ph.GA},
       adsurl = {https://ui.adsabs.harvard.edu/abs/2025ApJ...987..186F}
}

@ARTICLE{Sobral13,
       author = {{Sobral}, David and {Smail}, Ian and {Best}, Philip N. and {Geach}, James E. and {Matsuda}, Yuichi and {Stott}, John P. and {Cirasuolo}, Michele and {Kurk}, Jaron},
        title = "{A large H{\ensuremath{\alpha}} survey at z = 2.23, 1.47, 0.84 and 0.40: the 11 Gyr evolution of star-forming galaxies from HiZELS★}",
      journal = {\mnras},
         year = 2013,
        month = jan,
       volume = {428},
       number = {2},
        pages = {1128-1146},
          doi = {10.1093/mnras/sts096},
archivePrefix = {arXiv},
       eprint = {1202.3436},
 primaryClass = {astro-ph.CO},
       adsurl = {https://ui.adsabs.harvard.edu/abs/2013MNRAS.428.1128S}
}

@ARTICLE{Lin25,
       author = {{Lin}, Xiaojing and {Egami}, Eiichi and {Sun}, Fengwu and {Zhang}, Haowen and {Fan}, Xiaohui and {Helton}, Jakob M. and {Wang}, Feige and {Bunker}, Andrew J. and {Cai}, Zheng and {Eisenstein}, Daniel J. and {Jaffe}, Daniel T. and {Ji}, Zhiyuan and {Jin}, Xiangyu and {Pudoka}, Maria Anne and {Tacchella}, Sandro and {Tee}, Wei Leong and {Rinaldi}, Pierluigi and {Robertson}, Brant and {Sun}, Yang and {Willmer}, Christopher N.~A. and {Willott}, Chris and {Zhang}, Junyu and {Zhu}, Yongda},
        title = "{The Luminosity Function and Clustering of H{\ensuremath{\alpha}} Emitting Galaxies at z ≍ 4{\ensuremath{-}}6 from a Complete NIRCam Grism Redshift Survey}",
      journal = {\apj},
         year = 2026,
        month = feb,
       volume = {997},
       number = {2},
          eid = {207},
        pages = {207},
          doi = {10.3847/1538-4357/ae225e},
archivePrefix = {arXiv},
       eprint = {2504.08028},
 primaryClass = {astro-ph.GA},
       adsurl = {https://ui.adsabs.harvard.edu/abs/2026ApJ...997..207L}
}

@ARTICLE{Stanway18,
       author = {{Stanway}, E.~R. and {Eldridge}, J.~J.},
        title = "{Re-evaluating old stellar populations}",
      journal = {\mnras},
         year = 2018,
        month = sep,
       volume = {479},
       number = {1},
        pages = {75-93},
          doi = {10.1093/mnras/sty1353},
archivePrefix = {arXiv},
       eprint = {1805.08784},
 primaryClass = {astro-ph.GA},
       adsurl = {https://ui.adsabs.harvard.edu/abs/2018MNRAS.479...75S}
}

@ARTICLE{Naidu24,
       author = {{Naidu}, Rohan P. and {Matthee}, Jorryt and {Kramarenko}, Ivan and {Weibel}, Andrea and {Brammer}, Gabriel and {Oesch}, Pascal A. and {Lechner}, Peter and {Furtak}, Lukas J. and {Di Cesare}, Claudia and {Torralba}, Alberto and {Kotiwale}, Gauri and {Bezanson}, Rachel and {Bouwens}, Rychard J. and {Chandra}, Vedant and {Claeyssens}, Ad{\'e}la{\"\i}de and {Danhaive}, A. Lola and {Frebel}, Anna and {de Graaff}, Anna and {Greene}, Jenny E. and {Heintz}, Kasper E. and {Ji}, Alexander P. and {Kashino}, Daichi and {Katz}, Harley and {Labbe}, Ivo and {Leja}, Joel and {Li}, Yijia and {Maseda}, Michael V. and {Richard}, Johan and {Shivaei}, Irene and {Simcoe}, Robert A. and {Sobral}, David and {Suess}, Katherine A. and {Tacchella}, Sandro and {Williams}, Christina C.},
        title = "{All the Little Things in Abell 2744: $>$1000 Gravitationally Lensed Dwarf Galaxies at $z=0-9$ from JWST NIRCam Grism Spectroscopy}",
      journal = {arXiv e-prints},
         year = 2024,
        month = oct,
          eid = {arXiv:2410.01874},
        pages = {arXiv:2410.01874},
          doi = {10.48550/arXiv.2410.01874},
archivePrefix = {arXiv},
       eprint = {2410.01874},
 primaryClass = {astro-ph.GA},
       adsurl = {https://ui.adsabs.harvard.edu/abs/2024arXiv241001874N}
}

@ARTICLE{Matthee22,
       author = {{Matthee}, Jorryt and {Feltre}, Anna and {Maseda}, Michael and {Nanayakkara}, Themiya and {Boogaard}, Leindert and {Bacon}, Roland and {Verhamme}, Anne and {Leclercq}, Floriane and {Kusakabe}, Haruka and {Urrutia}, Tanya and {Wisotzki}, Lutz},
        title = "{Deciphering stellar metallicities in the early Universe: case study of a young galaxy at z = 4.77 in the MUSE eXtremely Deep Field}",
      journal = {\aap},
         year = 2022,
        month = apr,
       volume = {660},
          eid = {A10},
        pages = {A10},
          doi = {10.1051/0004-6361/202142187},
archivePrefix = {arXiv},
       eprint = {2111.14855},
 primaryClass = {astro-ph.GA},
       adsurl = {https://ui.adsabs.harvard.edu/abs/2022A&A...660A..10M}
}

@ARTICLE{Cullen19,
       author = {{Cullen}, F. and {McLure}, R.~J. and {Dunlop}, J.~S. and {Khochfar}, S. and {Dav{\'e}}, R. and {Amor{\'\i}n}, R. and {Bolzonella}, M. and {Carnall}, A.~C. and {Castellano}, M. and {Cimatti}, A. and {Cirasuolo}, M. and {Cresci}, G. and {Fynbo}, J.~P.~U. and {Fontanot}, F. and {Gargiulo}, A. and {Garilli}, B. and {Guaita}, L. and {Hathi}, N. and {Hibon}, P. and {Mannucci}, F. and {Marchi}, F. and {McLeod}, D.~J. and {Pentericci}, L. and {Pozzetti}, L. and {Shapley}, A.~E. and {Talia}, M. and {Zamorani}, G.},
        title = "{The VANDELS survey: the stellar metallicities of star-forming galaxies at 2.5 < z < 5.0}",
      journal = {\mnras},
         year = 2019,
        month = aug,
       volume = {487},
       number = {2},
        pages = {2038-2060},
          doi = {10.1093/mnras/stz1402},
archivePrefix = {arXiv},
       eprint = {1903.11081},
 primaryClass = {astro-ph.GA},
       adsurl = {https://ui.adsabs.harvard.edu/abs/2019MNRAS.487.2038C}
}

@ARTICLE{Scholte25,
       author = {{Scholte}, D. and {Cullen}, F. and {Carnall}, A.~C. and {Arellano-C{\'o}rdova}, K.~Z. and {Stanton}, T.~M. and {Barrufet}, L. and {Begley}, R. and {Bondestam}, C. and {Donnan}, C.~T. and {Dunlop}, J.~S. and {Leung}, H.-H. and {McLeod}, D.~J. and {McLure}, R.~J. and {Moustakas}, J.~M. and {Pollock}, C.~L. and {Shapley}, A.~E. and {Stevenson}, S. and {Zou}, H.},
        title = "{The JWST EXCELS survey: probing strong-line diagnostics and the chemical evolution of galaxies over cosmic time using T$_{e}$-metallicities}",
      journal = {\mnras},
         year = 2025,
        month = jun,
       volume = {540},
       number = {2},
        pages = {1800-1826},
          doi = {10.1093/mnras/staf834},
archivePrefix = {arXiv},
       eprint = {2502.10499},
 primaryClass = {astro-ph.GA},
       adsurl = {https://ui.adsabs.harvard.edu/abs/2025MNRAS.540.1800S}
}

@ARTICLE{Curti24,
       author = {{Curti}, Mirko and {Maiolino}, Roberto and {Curtis-Lake}, Emma and {Chevallard}, Jacopo and {Carniani}, Stefano and {D'Eugenio}, Francesco and {Looser}, Tobias J. and {Scholtz}, Jan and {Charlot}, Stephane and {Cameron}, Alex and {{\"U}bler}, Hannah and {Witstok}, Joris and {Boyett}, Kristian and {Laseter}, Isaac and {Sandles}, Lester and {Arribas}, Santiago and {Bunker}, Andrew and {Giardino}, Giovanna and {Maseda}, Michael V. and {Rawle}, Tim and {Rodr{\'\i}guez Del Pino}, Bruno and {Smit}, Renske and {Willott}, Chris J. and {Eisenstein}, Daniel J. and {Hausen}, Ryan and {Johnson}, Benjamin and {Rieke}, Marcia and {Robertson}, Brant and {Tacchella}, Sandro and {Williams}, Christina C. and {Willmer}, Christopher and {Baker}, William M. and {Bhatawdekar}, Rachana and {Egami}, Eiichi and {Helton}, Jakob M. and {Ji}, Zhiyuan and {Kumari}, Nimisha and {Perna}, Michele and {Shivaei}, Irene and {Sun}, Fengwu},
        title = "{JADES: Insights into the low-mass end of the mass-metallicity-SFR relation at 3 < z < 10 from deep JWST/NIRSpec spectroscopy}",
      journal = {\aap},
         year = 2024,
        month = apr,
       volume = {684},
          eid = {A75},
        pages = {A75},
          doi = {10.1051/0004-6361/202346698},
archivePrefix = {arXiv},
       eprint = {2304.08516},
 primaryClass = {astro-ph.GA},
       adsurl = {https://ui.adsabs.harvard.edu/abs/2024A&A...684A..75C}
}

@ARTICLE{Asada24,
       author = {{Asada}, Yoshihisa and {Sawicki}, Marcin and {Abraham}, Roberto and {Brada{\v{c}}}, Maru{\v{s}}a and {Brammer}, Gabriel and {Desprez}, Guillaume and {Estrada-Carpenter}, Vince and {Iyer}, Kartheik and {Martis}, Nicholas and {Matharu}, Jasleen and {Mowla}, Lamiya and {Muzzin}, Adam and {Noirot}, Ga{\"e}l and {Sarrouh}, Ghassan T.~E. and {Strait}, Victoria and {Willott}, Chris J. and {Harshan}, Anishya},
        title = "{Bursty star formation and galaxy-galaxy interactions in low-mass galaxies 1 Gyr after the Big Bang}",
      journal = {\mnras},
         year = 2024,
        month = feb,
       volume = {527},
       number = {4},
        pages = {11372-11392},
          doi = {10.1093/mnras/stad3902},
archivePrefix = {arXiv},
       eprint = {2310.02314},
 primaryClass = {astro-ph.GA},
       adsurl = {https://ui.adsabs.harvard.edu/abs/2024MNRAS.52711372A}
}

@ARTICLE{Cole25,
       author = {{Cole}, Justin W. and {Papovich}, Casey and {Finkelstein}, Steven L. and {Bagley}, Micaela B. and {Dickinson}, Mark and {Iyer}, Kartheik G. and {Yung}, L.~Y. Aaron and {Ciesla}, Laure and {Amor{\'\i}n}, Ricardo O. and {Arrabal Haro}, Pablo and {Bhatawdekar}, Rachana and {Calabr{\`o}}, Antonello and {Cleri}, Nikko J. and {de la Vega}, Alexander and {Dekel}, Avishai and {Endsley}, Ryan and {Gawiser}, Eric and {Giavalisco}, Mauro and {Hathi}, Nimish P. and {Hirschmann}, Michaela and {Holwerda}, Benne W. and {Kartaltepe}, Jeyhan S. and {Koekemoer}, Anton M. and {Lucas}, Ray A. and {Mascia}, Sara and {Mobasher}, Bahram and {P{\'e}rez-Gonz{\'a}lez}, Pablo G. and {Rodighiero}, Giulia and {Ronayne}, Kaila and {Tacchella}, Sandro and {Weiner}, Benjamin J. and {Wilkins}, Stephen M.},
        title = "{CEERS: Increasing Scatter along the Star-forming Main Sequence Indicates Early Galaxies Form in Bursts}",
      journal = {\apj},
         year = 2025,
        month = feb,
       volume = {979},
       number = {2},
          eid = {193},
        pages = {193},
          doi = {10.3847/1538-4357/ad9a6a},
archivePrefix = {arXiv},
       eprint = {2312.10152},
 primaryClass = {astro-ph.GA},
       adsurl = {https://ui.adsabs.harvard.edu/abs/2025ApJ...979..193C}
}

@ARTICLE{Endsley24,
       author = {{Endsley}, Ryan and {Stark}, Daniel P. and {Whitler}, Lily and {Topping}, Michael W. and {Johnson}, Benjamin D. and {Robertson}, Brant and {Tacchella}, Sandro and {Alberts}, Stacey and {Baker}, William M. and {Bhatawdekar}, Rachana and {Boyett}, Kristan and {Bunker}, Andrew J. and {Cameron}, Alex J. and {Carniani}, Stefano and {Charlot}, Stephane and {Chen}, Zuyi and {Chevallard}, Jacopo and {Curtis-Lake}, Emma and {Danhaive}, A. Lola and {Egami}, Eiichi and {Eisenstein}, Daniel J. and {Hainline}, Kevin and {Helton}, Jakob M. and {Ji}, Zhiyuan and {Looser}, Tobias J. and {Maiolino}, Roberto and {Nelson}, Erica and {Pusk{\'a}s}, D{\'a}vid and {Rieke}, George and {Rieke}, Marcia and {Rix}, Hans-Walter and {Sandles}, Lester and {Saxena}, Aayush and {Simmonds}, Charlotte and {Smit}, Renske and {Sun}, Fengwu and {Williams}, Christina C. and {Willmer}, Christopher N.~A. and {Willott}, Chris and {Witstok}, Joris},
        title = "{The star-forming and ionizing properties of dwarf z 6-9 galaxies in JADES: insights on bursty star formation and ionized bubble growth}",
      journal = {\mnras},
         year = 2024,
        month = sep,
       volume = {533},
       number = {1},
        pages = {1111-1142},
          doi = {10.1093/mnras/stae1857},
archivePrefix = {arXiv},
       eprint = {2306.05295},
 primaryClass = {astro-ph.GA},
       adsurl = {https://ui.adsabs.harvard.edu/abs/2024MNRAS.533.1111E}
}

@ARTICLE{Gotberg17,
       author = {{G{\"o}tberg}, Y. and {de Mink}, S.~E. and {Groh}, J.~H.},
        title = "{Ionizing spectra of stars that lose their envelope through interaction with a binary companion: role of metallicity}",
      journal = {\aap},
         year = 2017,
        month = nov,
       volume = {608},
          eid = {A11},
        pages = {A11},
          doi = {10.1051/0004-6361/201730472},
archivePrefix = {arXiv},
       eprint = {1701.07439},
 primaryClass = {astro-ph.SR},
       adsurl = {https://ui.adsabs.harvard.edu/abs/2017A&A...608A..11G}
}

@ARTICLE{Covelo-Paz25,
       author = {{Covelo-Paz}, Alba and {Giovinazzo}, Emma and {Oesch}, Pascal A. and {Meyer}, Romain A. and {Weibel}, Andrea and {Brammer}, Gabriel and {Fudamoto}, Yoshinobu and {Kerutt}, Josephine and {Lin}, Jamie and {Matharu}, Jasleen and {Naidu}, Rohan P. and {Velichko}, Anna and {Bollo}, Victoria and {Bouwens}, Rychard and {Chisholm}, John and {Illingworth}, Garth D. and {Kramarenko}, Ivan and {Magee}, Daniel and {Maseda}, Michael and {Matthee}, Jorryt and {Nelson}, Erica and {Reddy}, Naveen and {Schaerer}, Daniel and {Stefanon}, Mauro and {Xiao}, Mengyuan},
        title = "{An H{\ensuremath{\alpha}} view of galaxy buildup in the first 2 Gyr: Luminosity functions at z {\ensuremath{\sim}} 4‑6.5 from NIRCam/grism spectroscopy}",
      journal = {\aap},
         year = 2025,
        month = feb,
       volume = {694},
          eid = {A178},
        pages = {A178},
          doi = {10.1051/0004-6361/202452363},
archivePrefix = {arXiv},
       eprint = {2409.17241},
 primaryClass = {astro-ph.GA},
       adsurl = {https://ui.adsabs.harvard.edu/abs/2025A&A...694A.178C}
}

@ARTICLE{Pirie24,
       author = {{Pirie}, C.~A. and {Best}, P.~N. and {Duncan}, K.~J. and {McLeod}, D.~J. and {Cochrane}, R.~K. and {Clausen}, M. and {Dunlop}, J.~S. and {Flury}, S.~R. and {Geach}, J.~E. and {Hale}, C.~L. and {Ibar}, E. and {Kondapally}, R. and {Li}, Zefeng and {Matthee}, J. and {McLure}, R.~J. and {Ossa-Fuentes}, L. and {Patrick}, A.~L. and {Smail}, Ian and {Sobral}, D. and {Stephenson}, H.~M.~O. and {Stott}, J.~P. and {Swinbank}, A.~M.},
        title = "{The JWST Emission Line Survey (JELS): an untargeted search for H {\ensuremath{\alpha}} emission line galaxies at z > 6 and their physical properties}",
      journal = {\mnras},
         year = 2025,
        month = aug,
       volume = {541},
       number = {2},
        pages = {1348-1376},
          doi = {10.1093/mnras/staf1006},
archivePrefix = {arXiv},
       eprint = {2410.11808},
 primaryClass = {astro-ph.GA},
       adsurl = {https://ui.adsabs.harvard.edu/abs/2025MNRAS.541.1348P}
}

@ARTICLE{Sun25,
       author = {{Sun}, Fengwu and {Wang}, Feige and {Yang}, Jinyi and {Champagne}, Jaclyn B. and {Decarli}, Roberto and {Fan}, Xiaohui and {Ba{\~n}ados}, Eduardo and {Cai}, Zheng and {Colina}, Luis and {Egami}, Eiichi and {Hennawi}, Joseph F. and {Jin}, Xiangyu and {Jun}, Hyunsung D. and {Khusanova}, Yana and {Li}, Mingyu and {Li}, Zihao and {Lin}, Xiaojing and {Liu}, Weizhe and {Meyer}, Romain A. and {Pudoka}, Maria A. and {Rieke}, George H. and {Shen}, Yue and {Tee}, Wei Leong and {Venemans}, Bram and {Walter}, Fabian and {Wu}, Yunjing and {Zhang}, Huanian and {Zou}, Siwei},
        title = "{A SPectroscopic Survey of Biased Halos in the Reionization Era (ASPIRE): Spectroscopically Complete Census of Obscured Cosmic Star Formation Rate Density at z = 4{\textendash}6}",
      journal = {\apj},
         year = 2025,
        month = feb,
       volume = {980},
       number = {1},
          eid = {12},
        pages = {12},
          doi = {10.3847/1538-4357/ad9d0e},
archivePrefix = {arXiv},
       eprint = {2412.06894},
 primaryClass = {astro-ph.GA},
       adsurl = {https://ui.adsabs.harvard.edu/abs/2025ApJ...980...12S}
}

@ARTICLE{Mascia24,
       author = {{Mascia}, S. and {Pentericci}, L. and {Calabr{\`o}}, A. and {Santini}, P. and {Napolitano}, L. and {Arrabal Haro}, P. and {Castellano}, M. and {Dickinson}, M. and {Ocvirk}, P. and {Lewis}, J.~S.~W. and {Amor{\'\i}n}, R. and {Bagley}, M. and {Bhatawdekar}, R. and {Cleri}, N.~J. and {Costantin}, L. and {Dekel}, A. and {Finkelstein}, S.~L. and {Fontana}, A. and {Giavalisco}, M. and {Grogin}, N.~A. and {Hathi}, N.~P. and {Hirschmann}, M. and {Holwerda}, B.~W. and {Jung}, I. and {Kartaltepe}, J.~S. and {Koekemoer}, A.~M. and {Lucas}, R.~A. and {Papovich}, C. and {P{\'e}rez-Gonz{\'a}lez}, P.~G. and {Pirzkal}, N. and {Trump}, J.~R. and {Wilkins}, S.~M. and {Yung}, L.~Y.~A.},
        title = "{New insight on the nature of cosmic reionizers from the CEERS survey}",
      journal = {\aap},
         year = 2024,
        month = may,
       volume = {685},
          eid = {A3},
        pages = {A3},
          doi = {10.1051/0004-6361/202347884},
archivePrefix = {arXiv},
       eprint = {2309.02219},
 primaryClass = {astro-ph.GA},
       adsurl = {https://ui.adsabs.harvard.edu/abs/2024A&A...685A...3M}
}

@ARTICLE{Leitherer99,
       author = {{Leitherer}, Claus and {Schaerer}, Daniel and {Goldader}, Jeffrey D. and {Delgado}, Rosa M. Gonz{\'a}lez and {Robert}, Carmelle and {Kune}, Denis Foo and {de Mello}, Du{\'\i}lia F. and {Devost}, Daniel and {Heckman}, Timothy M.},
        title = "{Starburst99: Synthesis Models for Galaxies with Active Star Formation}",
      journal = {\apjs},
         year = 1999,
        month = jul,
       volume = {123},
       number = {1},
        pages = {3-40},
          doi = {10.1086/313233},
archivePrefix = {arXiv},
       eprint = {astro-ph/9902334},
 primaryClass = {astro-ph},
       adsurl = {https://ui.adsabs.harvard.edu/abs/1999ApJS..123....3L}
}

@ARTICLE{Bouwens15,
       author = {{Bouwens}, R.~J. and {Illingworth}, G.~D. and {Oesch}, P.~A. and {Trenti}, M. and {Labb{\'e}}, I. and {Bradley}, L. and {Carollo}, M. and {van Dokkum}, P.~G. and {Gonzalez}, V. and {Holwerda}, B. and {Franx}, M. and {Spitler}, L. and {Smit}, R. and {Magee}, D.},
        title = "{UV Luminosity Functions at Redshifts z {\ensuremath{\sim}} 4 to z {\ensuremath{\sim}} 10: 10,000 Galaxies from HST Legacy Fields}",
      journal = {\apj},
         year = 2015,
        month = apr,
       volume = {803},
       number = {1},
          eid = {34},
        pages = {34},
          doi = {10.1088/0004-637X/803/1/34},
archivePrefix = {arXiv},
       eprint = {1403.4295},
 primaryClass = {astro-ph.CO},
       adsurl = {https://ui.adsabs.harvard.edu/abs/2015ApJ...803...34B}
}

@ARTICLE{Zavala21,
       author = {{Zavala}, J.~A. and {Casey}, C.~M. and {Manning}, S.~M. and {Aravena}, M. and {Bethermin}, M. and {Caputi}, K.~I. and {Clements}, D.~L. and {Cunha}, E. da and {Drew}, P. and {Finkelstein}, S.~L. and {Fujimoto}, S. and {Hayward}, C. and {Hodge}, J. and {Kartaltepe}, J.~S. and {Knudsen}, K. and {Koekemoer}, A.~M. and {Long}, A.~S. and {Magdis}, G.~E. and {Man}, A.~W.~S. and {Popping}, G. and {Sanders}, D. and {Scoville}, N. and {Sheth}, K. and {Staguhn}, J. and {Toft}, S. and {Treister}, E. and {Vieira}, J.~D. and {Yun}, M.~S.},
        title = "{The Evolution of the IR Luminosity Function and Dust-obscured Star Formation over the Past 13 Billion Years}",
      journal = {\apj},
         year = 2021,
        month = mar,
       volume = {909},
       number = {2},
          eid = {165},
        pages = {165},
          doi = {10.3847/1538-4357/abdb27},
archivePrefix = {arXiv},
       eprint = {2101.04734},
 primaryClass = {astro-ph.GA},
       adsurl = {https://ui.adsabs.harvard.edu/abs/2021ApJ...909..165Z}
}

@ARTICLE{Davies19,
       author = {{Davies}, L.~J.~M. and {Lagos}, C. del P. and {Katsianis}, A. and {Robotham}, A.~S.~G. and {Cortese}, L. and {Driver}, S.~P. and {Bremer}, M.~N. and {Brown}, M.~J.~I. and {Brough}, S. and {Cluver}, M.~E. and {Grootes}, M.~W. and {Holwerda}, B.~W. and {Owers}, M. and {Phillipps}, S.},
        title = "{Galaxy And Mass Assembly (GAMA): The sSFR-M$_{*}$ relation part I - {\ensuremath{\sigma}}$_{sSFR}$-M$_{*}$ as a function of sample, SFR indicator, and environment}",
      journal = {\mnras},
         year = 2019,
        month = feb,
       volume = {483},
       number = {2},
        pages = {1881-1900},
          doi = {10.1093/mnras/sty2957},
archivePrefix = {arXiv},
       eprint = {1811.03712},
 primaryClass = {astro-ph.GA},
       adsurl = {https://ui.adsabs.harvard.edu/abs/2019MNRAS.483.1881D}
}

@ARTICLE{Tacchella20,
       author = {{Tacchella}, Sandro and {Forbes}, John C. and {Caplar}, Neven},
        title = "{Stochastic modelling of star-formation histories II: star-formation variability from molecular clouds and gas inflow}",
      journal = {\mnras},
         year = 2020,
        month = sep,
       volume = {497},
       number = {1},
        pages = {698-725},
          doi = {10.1093/mnras/staa1838},
archivePrefix = {arXiv},
       eprint = {2006.09382},
 primaryClass = {astro-ph.GA},
       adsurl = {https://ui.adsabs.harvard.edu/abs/2020MNRAS.497..698T}
}

@ARTICLE{Flores21,
       author = {{Flores Vel{\'a}zquez}, Jos{\'e} A. and {Gurvich}, Alexander B. and {Faucher-Gigu{\`e}re}, Claude-Andr{\'e} and {Bullock}, James S. and {Starkenburg}, Tjitske K. and {Moreno}, Jorge and {Lazar}, Alexandres and {Mercado}, Francisco J. and {Stern}, Jonathan and {Sparre}, Martin and {Hayward}, Christopher C. and {Wetzel}, Andrew and {El-Badry}, Kareem},
        title = "{The time-scales probed by star formation rate indicators for realistic, bursty star formation histories from the FIRE simulations}",
      journal = {\mnras},
         year = 2021,
        month = mar,
       volume = {501},
       number = {4},
        pages = {4812-4824},
          doi = {10.1093/mnras/staa3893},
archivePrefix = {arXiv},
       eprint = {2008.08582},
 primaryClass = {astro-ph.GA},
       adsurl = {https://ui.adsabs.harvard.edu/abs/2021MNRAS.501.4812F}
}

@ARTICLE{EWang20,
       author = {{Wang}, Enci and {Lilly}, Simon J.},
        title = "{The Variability of the Star Formation Rate in Galaxies. I. Star Formation Histories Traced by EW(H{\ensuremath{\alpha}}) and EW(H{\ensuremath{\delta}}$_{A}$)}",
      journal = {\apj},
         year = 2020,
        month = apr,
       volume = {892},
       number = {2},
          eid = {87},
        pages = {87},
          doi = {10.3847/1538-4357/ab7b7d},
archivePrefix = {arXiv},
       eprint = {1912.06523},
 primaryClass = {astro-ph.GA},
       adsurl = {https://ui.adsabs.harvard.edu/abs/2020ApJ...892...87W}
}

@ARTICLE{Iyer2020,
       author = {{Iyer}, Kartheik G. and {Tacchella}, Sandro and {Genel}, Shy and {Hayward}, Christopher C. and {Hernquist}, Lars and {Brooks}, Alyson M. and {Caplar}, Neven and {Dav{\'e}}, Romeel and {Diemer}, Benedikt and {Forbes}, John C. and {Gawiser}, Eric and {Somerville}, Rachel S. and {Starkenburg}, Tjitske K.},
        title = "{The diversity and variability of star formation histories in models of galaxy evolution}",
      journal = {\mnras},
         year = 2020,
        month = oct,
       volume = {498},
       number = {1},
        pages = {430-463},
          doi = {10.1093/mnras/staa2150},
archivePrefix = {arXiv},
       eprint = {2007.07916},
 primaryClass = {astro-ph.GA},
       adsurl = {https://ui.adsabs.harvard.edu/abs/2020MNRAS.498..430I}
}

@ARTICLE{MattheeSchaye19,
       author = {{Matthee}, Jorryt and {Schaye}, Joop},
        title = "{The origin of scatter in the star formation rate-stellar mass relation}",
      journal = {\mnras},
         year = 2019,
        month = mar,
       volume = {484},
       number = {1},
        pages = {915-932},
          doi = {10.1093/mnras/stz030},
archivePrefix = {arXiv},
       eprint = {1805.05956},
 primaryClass = {astro-ph.GA},
       adsurl = {https://ui.adsabs.harvard.edu/abs/2019MNRAS.484..915M}
}

@ARTICLE{Wan25,
       author = {{Wan}, Jenny T. and {Tacchella}, Sandro and {D'Eugenio}, Francesco and {Johnson}, Benjamin D. and {van der Wel}, Arjen},
        title = "{Decoding the variability in the star formation histories of z {\ensuremath{\sim}} 0.8 galaxies}",
      journal = {\mnras},
         year = 2025,
        month = jun,
       volume = {539},
       number = {4},
        pages = {2891-2909},
          doi = {10.1093/mnras/staf657},
archivePrefix = {arXiv},
       eprint = {2504.05281},
 primaryClass = {astro-ph.GA},
       adsurl = {https://ui.adsabs.harvard.edu/abs/2025MNRAS.539.2891W}
}

@ARTICLE{Sparre15,
       author = {{Sparre}, Martin and {Hayward}, Christopher C. and {Springel}, Volker and {Vogelsberger}, Mark and {Genel}, Shy and {Torrey}, Paul and {Nelson}, Dylan and {Sijacki}, Debora and {Hernquist}, Lars},
        title = "{The star formation main sequence and stellar mass assembly of galaxies in the Illustris simulation}",
      journal = {\mnras},
         year = 2015,
        month = mar,
       volume = {447},
       number = {4},
        pages = {3548-3563},
          doi = {10.1093/mnras/stu2713},
archivePrefix = {arXiv},
       eprint = {1409.0009},
 primaryClass = {astro-ph.GA},
       adsurl = {https://ui.adsabs.harvard.edu/abs/2015MNRAS.447.3548S}
}

@ARTICLE{Bouche10,
       author = {{Bouch{\'e}}, N. and {Dekel}, A. and {Genzel}, R. and {Genel}, S. and {Cresci}, G. and {F{\"o}rster Schreiber}, N.~M. and {Shapiro}, K.~L. and {Davies}, R.~I. and {Tacconi}, L.},
        title = "{The Impact of Cold Gas Accretion Above a Mass Floor on Galaxy Scaling Relations}",
      journal = {\apj},
         year = 2010,
        month = aug,
       volume = {718},
       number = {2},
        pages = {1001-1018},
          doi = {10.1088/0004-637X/718/2/1001},
archivePrefix = {arXiv},
       eprint = {0912.1858},
 primaryClass = {astro-ph.CO},
       adsurl = {https://ui.adsabs.harvard.edu/abs/2010ApJ...718.1001B}
}

@ARTICLE{Kroupa2001,
       author = {{Kroupa}, Pavel},
        title = "{On the variation of the initial mass function}",
      journal = {\mnras},
         year = 2001,
        month = apr,
       volume = {322},
       number = {2},
        pages = {231-246},
          doi = {10.1046/j.1365-8711.2001.04022.x},
archivePrefix = {arXiv},
       eprint = {astro-ph/0009005},
 primaryClass = {astro-ph},
       adsurl = {https://ui.adsabs.harvard.edu/abs/2001MNRAS.322..231K}
}

\begin{appendix}
    \section{Timescales of star formation traced by \ha{}}
    \label{sec:sf-timescales}

\begin{figure}
        \centering
        \includegraphics[width=\linewidth]{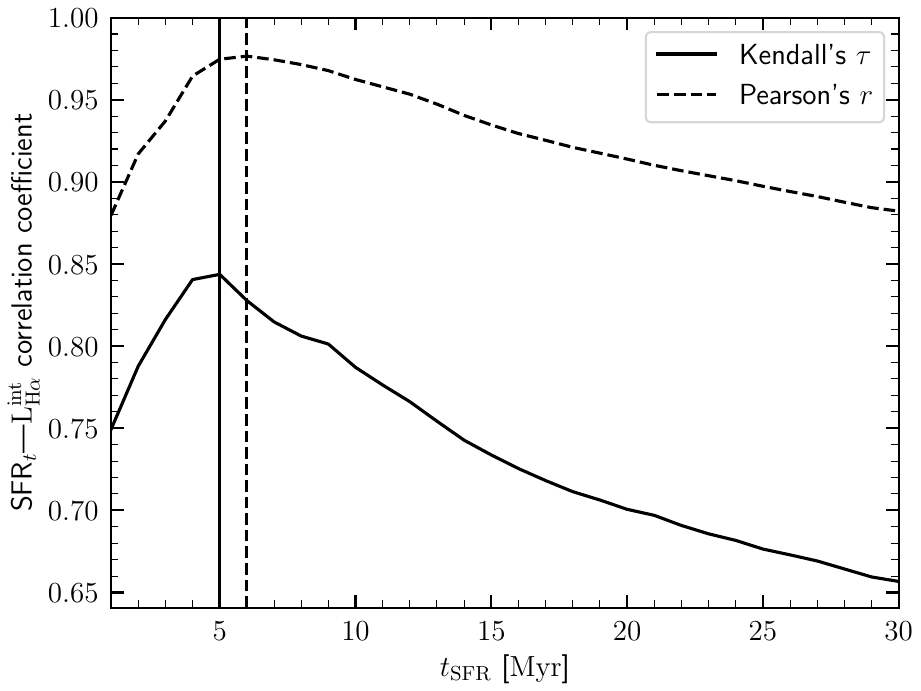}
        \caption{Correlation coefficients between the SFR and \lumhaint{} as a function of the star formation timescale, \tsfr{}.}
        \label{fig:t-sfr-best}
    \end{figure}

    It is commonly assumed in the literature that \ha{} traces star formation over the last $\tsfr{} = 10$~Myr. To test this, we use the SFHs published as part of SPDRv1 and calculate the average of the SFR over the last $\tsfr{}=1,2,\ldots,30$~Myr. We then calculate the correlation coefficients between \lumhaint{} and the SFRs averaged over different timescales. We find that Pearson's $r$ is highest between $4$~$\mathrm{Myr}\lesssim t \lesssim 12$~Myr ($r > 0.95$), with a maximum at $\tsfr{} = 6$~Myr ($r = 0.98$; \cref{fig:t-sfr-best}, dashed line). Similarly, Kendall's $\tau$ peaks at $\tsfr{} = 5$~Myr ($\tau = 0.84$; \cref{fig:t-sfr-best}, solid line). This suggests that a slightly shorter timescale ($\tsfr{} = 5$--$6$~Myr) might be more appropriate for \ha{}. Nevertheless, for consistency with previous works, we use the SFRs averaged over $\tsfr{}=10$~Myr throughout this paper.

    \section{\sfrmodel{} calibrations for the dust-attenuated \ha{} line measurements}
    \label{sec:sfr-calib-dust}

    \begin{equation}
        \label{eq:sfr-calib-lha-obs-corr}
        \begin{split}
        \logsfrfrac{} =& \loglumhafracobs{} - 41.00  \\
                       & + 0.66 \left( \loglumhafracobs{} - 41.46 \right)
        \end{split}
    \end{equation}

    \begin{equation}
        \label{eq:sfr-calib-ewha-obs-corr}
        \begin{split}
        \logsfrfrac{} =&  \loglumhafracobs{} - 41.00  \\
                       & + 0.23 \left( \loglumhafracobs{} - 41.46 \right)  \\
                       & - 0.67 \left( \logewhafracobs{} - 2.41 \right)
        \end{split}
    \end{equation}

    \begin{equation}
        \label{eq:sfr-calib-bd-corr}
        \begin{split}
        \logsfrfrac{} =& \loglumhafracobs{} - 41.00  \\
                       & + 0.58 \left( \loglumhafracobs{} - 41.46 \right)  \\
                       & + 5.81 \left( \logbd{} - 0.51 \right)
        \end{split}
    \end{equation}

    \begin{equation}
        \label{eq:sfr-calib-ewha-obs-bd-corr}
        \begin{split}
        \logsfrfrac{} =&  \loglumhafracobs{} - 41.00  \\
                       & + 0.22 \left( \loglumhafracobs{} - 41.46 \right)  \\
                       & - 0.59 \left( \logewhafracobs{} - 2.41 \right)  \\
                       & + 3.74 \left( \logbd{} - 0.51 \right)
        \end{split}
    \end{equation}

\end{appendix}

\end{document}